\documentclass[12pt,twoside]{article}
\usepackage[scale={.83,.83}]{geometry}
\usepackage[latin1]{inputenc}
\usepackage{amsmath,amssymb,bbm,mathrsfs,slashed}
\usepackage{axodraw,wrapfig,caption2,subfigure}
\usepackage[dvips]{epsfig}
\usepackage{hyperref}

\newcommand{\EM}{\text{\sf EM}} 
\newcommand{\ew}{\text{\sf EW}} 
\newcommand{\gtr}{\mbox{${\sf SU(3)_C \times SU(3)_L \times SU(3)_R
      \times \mathbbm{Z}_3}$}}  
\newcommand{\gtrs}{\mbox{${\sf [SU(3)]^3 \times \mathbbm{Z}_3}$}}
\newcommand{\hi}{\text{\sc h}}  
\newcommand{\Le}{\text{\sc l}}  
\newcommand{\qu}{\text{\sc q}}  
\newcommand{\VEV}[1]{\left\langle #1\right\rangle}        

\begin{document}

\title{ {\normalsize January 2006 \hfill \phantom{end}
  }\\
  \vspace{50pt} %
  \textbf{Minimal Trinification} %
  \\[30pt]}

\author{J. Sayre, S.~Wiesenfeldt and S. Willenbrock
  \\[20pt]
  {\normalsize {\slshape
      \begin{minipage}{.8\linewidth}
        \centering
        Department of Physics, University of Illinois at
        Urbana-Champaign
        \\[3pt]
        1110 West Green Street, Urbana, IL 61801, USA
      \end{minipage}
    }}
    \\[20pt]
  }

\date{\mbox{\phantom{date}}}

\maketitle

\thispagestyle{empty}

\begin{abstract}
  \noindent
  We study the trinified model, $\gtr$, with the minimal Higgs sector
  required for symmetry breaking.  There are five Higgs doublets, and
  gauge-coupling unification results if all five are at the weak
  scale, without supersymmetry.  The radiative see-saw mechanism
  yields sub-eV neutrino masses, without the need for intermediate
  scales, additional Higgs fields, or higher-dimensional operators.
  The proton lifetime is above the experimental limits, with the decay
  modes $p\to\bar\nu K^+$ and $p\to\mu^+ K^0$ potentially observable.
  We also consider supersymmetric versions of the model, with one or
  two Higgs doublets at the weak scale.  The radiative see-saw
  mechanism fails with weak-scale supersymmetry due to the
  nonrenormalization of the superpotential, but operates in the
  split-SUSY scenario.
\end{abstract}

\newpage


\section{Introduction \label{se:intro}}

Grand unification of the strong, weak, and electromagnetic
interactions into a simple gauge group is a very appealing idea that
has been vigorously pursued for many years \cite{Georgi:1974sy}. In
recent years most effort has gone into {\sffamily SU(5)} and
{\sffamily SO(10)} grand-unified models, as well as ${\sf E_6}$. Among
its most notable successes, grand unification predicted neutrino
masses in the $10^{-5}-10^2$ eV range \cite{Langacker:1980js},
compatible with the masses deduced from neutrino oscillation
experiments \cite{nu-masses}.

An alternative to unification into a simple gauge group is unification
into a product group, with identical factor groups.  The simplest and
most promising theory is the trinified model, ${\sf [SU(3)]^3=SU(3)_C
  \times SU(3)_L \times SU(3)_R}$
\cite{Achiman:1978rv,Glashow:1984gc}.  The equality of the three gauge
couplings is enforced by a discrete symmetry, such as $\mathbbm{Z}_3$.
Alternatively, the equality of the gauge couplings may be a result of
some additional structure at the unification scale, such as string
theory \cite{Ginsparg:1987ee}.

${\sf [SU(3)]}^3$ is a subgroup of ${\sf E_6}$, as are {\sffamily
  SU(5)} and {\sffamily SO(10)}.  All of these groups share the common
feature that $\sin^2\theta_W=\frac{3}{8}$ at the grand-unified scale.
The ${\sf [SU(3)]}^3$ model, however, is less unified than the models
based on simple groups.  This is both its weakness and its strength:
it makes fewer predictions, but it also does not run into
phenomenological difficulties as readily.

The implications of (non-supersymmetric) trinification were first
studied in some detail in Ref.~\cite{Babu:1985gi}, but with restricted
Yukawa couplings to leptons.\footnote{The supersymmetric model has
  been studied in Refs.~\cite{susy-su3}, recently also in the context
  of orbifold GUTs \cite{orbifold-su3}.}  Those couplings not only
determine the lepton masses but also the couplings of colored Higgs
bosons to quarks, which yield proton decay.  In this paper we revisit
this model, and completely elucidate its structure without making any
restrictions on the Yukawa couplings.  Furthermore,
Ref.~\cite{Babu:1985gi} used a crude technique to calculate the proton
decay branching ratios.  We will show that with unrestricted lepton
Yukawa couplings and a better technique for the calculation of proton
decay, the branching ratios are significantly different than those
previously obtained.

Trinification has a number of nice features beyond those of models
based on simple gauge groups:
\begin{itemize}
\item Baryon number is conserved by the gauge interactions.  This
  allows the possibility of lowering the grand-unified scale.  The
  (non-supersymmetric) standard model with six Higgs doublets yields
  gauge-coupling unification (with $\sin^2\theta_W=\frac{3}{8}$) at
  $M_\text{U}\simeq 10^{14}$ GeV \cite{Willenbrock:2003ca}.  As we
  shall see, the minimal\footnote{By minimal we mean a Higgs sector
    that contains just the fields needed to break ${\sf [SU(3)]}^3$ to
    the standard model.} trinified model can have as many as five
  light Higgs doublets, which is also sufficient for gauge-coupling
  unification.  As usual, the supersymmetric version of the theory
  needs just two light Higgs doublets (and their superpartners), or
  even just one in the split-SUSY scenario \cite{split-susy}.
  
\item In the minimal model, the only Higgs representations needed to
  break the gauge group to the standard model and to generate
  realistic quark and lepton masses are built from the defining
  representations of ${\sf SU(3)}$.  In particular, no adjoint Higgs
  field is needed.  Furthermore, the light Higgs doublets that break
  the electroweak symmetry lie in the same representations as the
  Higgs field that break the grand-unified symmetry.
  
\item The minimal renormalizable model is sufficiently flexible to
  accommodate any quark and lepton masses and mixing angles.  This may
  be viewed as a strength or a weakness, as mentioned above.
  Neutrinos acquire eV-scale masses via a ``radiative see-saw''
  mechanism \cite{Witten:1979nr}, both in the non-supersymmetric and
  the split-SUSY versions of the model.  If the mass differences of
  the SUSY partners is \mbox{${\cal O}\left(1\,\text{TeV}\right)$},
  this mechanism fails.
\end{itemize}

As mentioned above, ${\sf [SU(3)]}^3$ is a subgroup of ${\sf E_6}$,
and thus also a subgroup of ${\sf E_8 \times E_8}$, which makes it a
candidate for embedding in the heterotic string.  However, ${\sf
  [SU(3)]}^3$ is a candidate for heterotic string constructions
itself, since adjoint fields are not needed to break this gauge group
to the standard model \cite{susy-su3,orbifold-su3}.  In addition, the
smallest ${\sf E_6}$ Higgs representation capable of breaking ${\sf
  E_6}$ to ${\sf [SU(3)]}^3$ is the {\sffamily 650}.  In this paper we
study the ${\sf [SU(3)]}^3$ model on its own.

This paper is organized as follows.  In Section~\ref{se:model} we
review the trinified model and derive the fermion masses and
eigenstates for one generation.  In Section~\ref{se:3gen} we consider
the case of three generations, in particular quark mixing and the
hierarchy of neutrino masses.  Section~\ref{se:pd} is devoted to
proton decay, where we estimate the dominant decay modes.  Conclusions
are given in Section~\ref{se:conclusion}.  Additional details are
given in several Appendices.


\section{Trinified Model \label{se:model}}

We begin by briefly reviewing the $\gtr$ model
\cite{Achiman:1978rv,Glashow:1984gc,Babu:1985gi}.\footnote{Instead of
  $\mathbbm{Z}_3$, Refs.~\cite{su3-s3} use $S_3$.}  The
$\mathbbm{Z}_3$ symmetry guarantees that the three gauge couplings are
equal at and above the grand-unified scale.  The gauge bosons are
assigned to the adjoint representation, the fermions to $\psi_\Le
\oplus \psi_{\qu^c} \oplus \psi_\qu \equiv \left(1,3,3^\ast\right)
\oplus \left(3^\ast,1,3\right) \oplus \left(3,3^\ast,1\right)$.  The
fermion multiplets decompose with respect to the standard model as
follows:\footnote{Our notation differs from that in
  Refs.~\cite{Achiman:1978rv,Glashow:1984gc,Babu:1985gi,susy-su3,orbifold-su3},
  since $\psi_\Le$ is an ${\sf SU(3)_L}$-triplet, which contains ${\sf
    SU(2)_L}$-doublets, whereas $\psi_\qu$ is an ${\sf
    SU(3)_L}$-\emph{anti}-triplet.  In the other notation, $\psi_\Le$
  contains ${\sf SU(2)_L}$-\emph{anti}-doublets while $\psi_\qu$
  contains ${\sf SU(2)_L}$-doublets.  The resulting fermion masses and
  mixings agree, however.}
\begin{subequations} \label{eq:field-decomposition}
  \begin{align}     \label{eq:field-decomposition-l}
    \psi_\Le & \to \left(1,2,\tfrac{1}{2}\right) \oplus 2
    \left(1,2,-\tfrac{1}{2}\right) \oplus \left(1,1,1\right) \oplus 2
    \left(1,1,0\right) ,
    \\
    \label{eq:field-decomposition-qbar}
    \psi_{\qu^c} & \to \left(3^\ast,1,-\tfrac{2}{3}\right) \oplus 2
    \left(3^\ast,1,\tfrac{1}{3}\right) ,
    \\
    \label{eq:field-decomposition-q}
    \psi_\qu & \to \left(3,2^\ast,\tfrac{1}{6}\right) \oplus
    \left(3,1,-\tfrac{1}{3}\right) .
  \end{align}
\end{subequations}
Thus we find the fifteen left-chiral fermions of the standard model
plus twelve additional fermions. More explicitly,
\begin{align} \label{eq:fermions}
  \psi_\Le & =
  \begin{pmatrix}
    \left(\mathscr{E}\right) & \left(E^c\right) & \left(
      \mathscr{L} \right) \cr \mathscr{N}_1 & e^c & \mathscr{N}_2
  \end{pmatrix}
  , & \psi_{\qu^c} & =
  \begin{pmatrix}
    \mathscr{D}^c \cr u^c \cr \mathscr{B}^c
  \end{pmatrix}
  , &
  \psi_\qu & = \left( \left( -d\quad u \right)\ B
    \vphantom{\frac{B}{B}} \right) .
\end{align}
The field $(-d,u)$ is the (conjugate of the) usual quark doublet
$Q=\left({u\atop d}\right)$, while $B$ is an additional color-triplet,
weak-singlet quark.  The field $u^c$ is the usual up-conjugate quark
field, while $\mathscr{D}^c$ and $\mathscr{B}^c$ have the quantum
numbers of the down-conjugate quark field.  The actual down-conjugate
field $d^c$ is a linear combination of the two, as we shall see. The
field $e^c$ is the usual positron field, and the lepton doublet is a
linear combination of $\mathscr{L}$ and $\mathscr{E}$.  The field
$E^c$ denotes a lepton doublet with the opposite hypercharge, and
$\mathscr{N}_1$ and $\mathscr{N}_2$ are sterile (with respect to the
standard model) fermions.

The gauge interactions have an accidental ${\sf U(1)_Q \times
  U(1)_{Q^c} \times U(1)_L}$ global symmetry corresponding to phase
rotations of the fermion multiplets $\psi_\qu,
\psi_{\qu^c},\psi_\Le$.\footnote{With three generations, the
  accidental global symmetry is ${\sf U(3)_Q \times U(3)_{Q^c} \times
    U(3)_L}$.  However, only the ${\sf U(1)}$ subgroups will be
  important for later discussion.}  The linear combination ${\sf
  U(1)_{Q-Q^c}}$ is proportional to baryon number, so proton decay is
not mediated by gauge interactions.  As we shall see, proton decay is
mediated by Yukawa interactions. The global symmetry ${\sf U(1)_L}$ is
not lepton number, as the lepton multiplet $\psi_\Le$ contains both
leptons and antileptons.

$\gtrs$ is broken by a pair of $\left(1,3,3^\ast\right)$ Higgs fields,
which we denote by $\Phi^{1,2}_\Le$,
\begin{align} \label{eq:higgs}
  \Phi_\Le^a & =
  \begin{pmatrix}
    \left(\phi_1^a\right) & \left(\phi_2^a\right) &
    \left(\phi_3^a\right) \cr S_1^a & S_2^a & S_3^a
  \end{pmatrix}
  , & \VEV{\Phi^1_\Le} & =
  \begin{pmatrix}
    u_1 & 0 & 0 \cr 0 & u_2 & 0 \cr 0 & 0 & v_1
  \end{pmatrix}
  , & \VEV{\Phi^2_\Le} & =
  \begin{pmatrix}
    n_1 & 0 & n_3 \cr 0 & n_2 & 0 \cr v_2 & 0 & v_3
  \end{pmatrix}
  ,
\end{align}
with $v_i={\cal O}\left(M_\text{U}\right)$ and $u_i,\, n_i={\cal
  O}\left(M_\ew\right)$, where $M_\text{U}$ and $M_\ew$ are the
unification and electroweak scales, respectively.  Both $v_1$ ($v_3$)
and $v_2$ break ${\sf SU(3)_L \times SU(3)_R}$ to \mbox{${\sf SU(2)_L
    \times SU(2)_R \times U(1)}$}, but the ${\sf SU(2)_R \times U(1)}$
are different.  Together they break $\gtrs$ to \mbox{${\sf SU(3)_C
    \times SU(2)_L \times U(1)_Y}$}.

Of the six Higgs doublets ($\phi_i$, three each in $\Phi^{1,2}_\Le$),
one linear combination\footnote{\mbox{$G\propto v_1\phi_3^1 +
    v_2\phi_1^2 + v_3\phi_3^2$}.}  is eaten by the gauge bosons that
acquire unification-scale masses. If the remaining five doublets have
electroweak-scale masses, then gauge-coupling unification results at
$M_\text{U}\simeq 10^{14}$ GeV without supersymmetry
\cite{Willenbrock:2003ca}.  In general it would take several
fine-tunings to arrange this, so it is an even more acute form of the
usual hierarchy problem.  We do not address this problem further in
this paper.

Another potential drawback of five light Higgs doublets is
Higgs-mediated flavor-changing neutral currents, which are present
whenever a fermion of a given electric charge couples to more than one
Higgs field \cite{Glashow:1976nt}.  Whether such interactions are
present at an acceptable level depends on the details of fermion mass
generation, which we do not pursue in this paper.  Higgs-mediated
flavor-changing neutral currents may be suppressed by small Yukawa
couplings \cite{fcnc}.

Alternatively, one may consider the supersymmetric version of the
theory, in which case just two light Higgs doublets (and their
superpartners) are required to yield successful gauge-coupling
unification \cite{unification}.  However, the supersymmetric version
of trinification, in its minimal incarnation, does not provide a
see-saw mechanism for neutrino masses.  This problem is ameliorated in
the split-SUSY version of the model \cite{split-susy}, as we will
discuss.

The electroweak symmetry is broken to ${\sf U(1)}_\EM$ when any of the
five Higgs doublets acquires an electroweak-scale vev.  These are
indicated by the vevs $u_i, n_i$ in Eq.~(\ref{eq:higgs}).

For simplicity, we set $v_3=0$ henceforth; this does not have any
effect on the qualitative aspects of the model.  In order to generate
masses for up-type quarks, we need $u_2$ and/or $n_2$ nonzero; for
down-type quarks and charged leptons, $u_1,n_1,n_3$ are the relevant
vevs.  Henceforth we choose $u_1$ and $u_2$ nonzero, and set $n_i =
0$.  Again, this does not affect the qualitative aspects of the model.
The general expressions for the fermion masses, with all vevs nonzero,
are given in Appendix~\ref{se:ferm-gen}.

\subsection{Yukawa Interactions \label{se:yukawa}}

With the Higgs fields $\Phi^a_\Le\left(1,3,3^\ast\right)$ (plus cyclic
permutations), two types of Yukawa couplings are allowed, namely
\mbox{$\psi_{\qu^c} \psi_\qu \Phi^a_\Le \equiv
  \left(\psi_{\qu^c}\right)^i_j \left(\psi_\qu\right)^j_k
  \left(\Phi^a_\Le\right)^k_i$} for the quarks and \mbox{$\psi_\Le
  \psi_\Le \Phi^a_\Le \equiv \epsilon^{ijk} \epsilon_{rst}
  \left(\psi_\Le\right)^r_i \left(\psi_\Le\right)^s_j
  \left(\Phi^a_\Le\right)^t_k$} for the leptons.  The former read
\begin{subequations} \label{eq:yukawa-quark}
  \begin{align} \label{eq:yukawa-quark-a}
    \mathscr{L}_q & = \psi_{\qu^c} \psi_\qu \left( g_1\, \Phi^1_\Le +
      g_2\, \Phi^2_\Le \right) + \text{cyclic} + \text{h.c.},
    \intertext{where}
    \psi_{\qu^c} \psi_\qu \Phi_\Le & = \mathscr{D}^c Q \phi_1 + u^c Q
    \phi_2 + \mathscr{B}^c Q \phi_3 + \mathscr{D}^c B S_1 + u^c B S_2
    + \mathscr{B}^c B S_3 \ .
    \label{eq:yukawa-quark-b}
  \end{align}
\end{subequations}
In Eq.~(\ref{eq:yukawa-quark-b}) we suppress both the superscript
on the Higgs field and $\epsilon=i\sigma_2$, which is implicit
between two ${\sf SU(2)_L}$ doublets here and throughout the
paper.  When $S_3^1$ and $S_1^2$ acquire the vevs $v_1$ and $v_2$,
respectively (see Eq.~(\ref{eq:higgs})), $B$ pairs up with a
linear combination of $\mathscr{D}^c$ and $\mathscr{B}^c$ to form
a Dirac fermion with a mass at the unification scale,
\begin{align} \label{eq:b-mass}
  m_B  & = \sqrt{g_1^2 v_1^2 + g_2^2 v_2^2}  \ .
\end{align}
The mass eigenstates are
\begin{align} \label{eq:down-es}
  d^c & = -s_\alpha\, \mathscr{D}^c + c_\alpha\, \mathscr{B}^c , & B^c
  & = c_\alpha\, \mathscr{D}^c + s_\alpha\, \mathscr{B}^c , &
  \tan\alpha & = \tfrac{g_1 v_1}{g_2 v_2} \ ,
\end{align}
where $s\equiv\sin,\ c\equiv\cos$.  We can express the Yukawa
couplings of these fields to the ${\sf SU(2)_L}$-doublet Higgs
fields as
\begin{align} \label{eq:yukawa-quark3}
  \mathscr{L}_q & = m_B\, B^c\, B \ + \sum_{a=1}^2 g_a\left[\left(
      -s_\alpha\, d^c + c_\alpha\, B^c \right) Q \phi_1^a + u^c Q
    \phi_2^a + \left( c_\alpha\, d^c + s_\alpha\, B^c \right) Q
    \phi_3^a\right] + \text{h.c.}
\end{align}
When $\phi^1_{1,2}$ acquire the vevs $u_{1,2}$, the light quarks
acquire masses
\begin{align} \label{eq:quarkmasses}
  m_u & = g_1\, u_2 \ , &
  m_d & = g_1\, u_1\, s_\alpha \ .
\end{align}
The heavy $B$ quark mass, as well as the quark mass eigenstates of
Eq.~(\ref{eq:down-es}), obtain tiny corrections ${\cal
  O}\left(\frac{u}{v}\right)$, which we neglect here and throughout
the paper.


The Yukawa couplings for the leptons are
\begin{align} \label{eq:yukawa-lepton}
  \mathscr{L}_l & = \tfrac{1}{2}\, \psi_\Le \psi_\Le \left( h_1\,
    \Phi^1_\Le + h_2\, \Phi^2_\Le \right) + \text{cyclic} +
  \text{h.c.} ,
  \intertext{where}
  \begin{split} \label{eq:yukawa-lepton2}
    \tfrac{1}{2}\, \psi_\Le \psi_\Le \Phi_\Le & = - \left( E^c
      \mathscr{N}_2 - \mathscr{L}\, e^c \right) \phi_1 + \left(
      \mathscr{E} \mathscr{N}_2 - \mathscr{L}\, \mathscr{N}_1 \right)
    \phi_2 + \left( E^c \mathscr{N}_1 - \mathscr{E} e^c \right) \phi_3
    \\
    & \mspace{21mu} + E^c \mathscr{L} S_1 - \mathscr{E} \mathscr{L} S_2
    - E^c \mathscr{E} S_3 \ .
  \end{split}
\end{align}
Only the first term in Eq.~(\ref{eq:yukawa-lepton}) is used in
Ref.~\cite{Babu:1985gi}, which is the restriction on the leptonic
Yukawa couplings mentioned in the Introduction. The doublets
$\mathscr{E}$ and $\mathscr{L}$ as well as the singlets
$\mathscr{N}_1$ and $\mathscr{N}_2$ mix to
{\allowdisplaybreaks
  \begin{subequations} \label{eq:lepton-es}
    \begin{align} \label{eq:doublet-es}
      E & = -s_\beta\ \mathscr{E}\ + c_\beta\, \mathscr{L} \ , &
      L & = \phantom{-}c_\beta\ \mathscr{E}\ + s_\beta\, \mathscr{L} \
      , & \tan\beta & = \tfrac{h_1 v_1}{h_2 v_2} \ ,
      \\
      \label{eq:n12-es}
      N_1 & = \phantom{-}s_\beta\, \mathscr{N}_1 - c_\beta\,
      \mathscr{N}_2 \ , &
      N_2 & = -c_\beta\, \mathscr{N}_1 - s_\beta\, \mathscr{N}_2 \ ,
    \end{align}
  \end{subequations}
}
so that
\begin{multline} \label{eq:yukawa-lepton3}
  \mathscr{L}_l = - m_E\, E^c E +\sum_{a=1}^2 h_i \left\{ - \left[ E^c
      \left( -c_\beta\, N_1 - s_\beta\, N_2 \right) - \left( c_\beta\,
        E + s_\beta\, L \right) e^c \right] \phi_1^a + \left( E N_2 -
      L N_1 \right) \phi_2^a \vphantom{\frac11} \right.
  \\[-2mm]
  + \left. \left[ E^c \left( s_\beta\, N_1 - c_\beta\, N_2 \right) -
      \left( - s_\beta\, E + c_\beta\, L \right) e^c\right] \phi_3^a
    \vphantom{\frac11} \right\} + \text{h.c.}
\end{multline}
The masses of the leptons are given by
\begin{align} \label{eq:lepton-masses}
  m_E & = \sqrt{h_1^2 v_1^2 + h_2^2 v_2^2} \ , &
  m_e & = h_1 u_1\, s_\beta \ , &
  m_{\nu, N_1} & = h_1 u_2 \ , &
  m_{N_2} & \simeq \frac{h_1^2 u_1 u_2\, s_\beta}{m_E} \ .
\end{align}
The general formulae for fermion masses, with all vevs nonzero,
are given in Appendix~\ref{se:ferm-gen}.

These results for the fermion masses show that, even in the minimal
model, there is no relation between the masses of the quarks and
leptons, since they depend on five independent parameters
($g_1,h_1,\frac{u_1}{u_2},s_\alpha,s_\beta$).  This is in contrast to
minimal $\sf SU(5)$, which yields $m_d=m_e$, or minimal ${\sf
  SO(10)}$, which yields $m_d=m_e$ and $m_u=m_\nu$.  Thus the minimal
trinification model is sufficient to describe the masses of the quarks
and charged leptons.  The additional matter charged under the standard
model group is vectorlike and superheavy; at tree level, however, the
model yields an active Dirac neutrino at the electroweak scale and a
sterile Majorana neutrino ($N_2$) at the eV scale, both in (potential)
conflict with observation.  As we shall see, this may be corrected at
one loop, via the ``radiative see-saw'' mechanism.

It is useful to understand the electroweak-scale Dirac neutrino
$\nu,N_1$, and the eV-scale sterile neutrino $N_2$, in terms of
accidental global symmetries. The accidental ${\sf U(1)_Q \times
  U(1)_{Q^c} \times U(1)_L}$ global symmetry of the gauge interactions
is violated by the Yukawa interactions, except for a ${\sf U(1)_X}$
subgroup in which the fermion multiplets carry charge $\frac{1}{2}$
and the Higgs fields carry charge $-1$.  This symmetry is broken when
the Higgs fields acquire unification-scale vevs $v_i$, but there
remains an unbroken global ${\sf U(1)}$ which is a linear combination
of ${\sf U(1)_X}$ and a broken gauge symmetry.  This unbroken global
symmetry is then broken by the electroweak-scale vev $u_2$, but again
an unbroken global ${\sf U(1)}$ survives. The fields $\nu,N_1$ pair up
to form a Dirac neutrino with a mass proportional to $u_2$ because
they have equal and opposite charges under this global ${\sf U(1)}$,
while the field $N_2$, which is also charged, remains massless. If
both $u_1$ and $u_2$ are nonzero, then no global symmetry survives.
This explains why the sterile neutrino $N_2$ acquires a Majorana mass
proportional to $u_1u_2$.\footnote{The formula for $m_{N_2}$ corrects
  Eq.~(4.14b) of Ref.~\cite{Babu:1985gi}.}  A detailed discussion of
the accidental global symmetries is contained in
Appendix~\ref{se:global}.


\subsection{One-loop Corrections to Neutrino Masses \label{se:neutrino}}

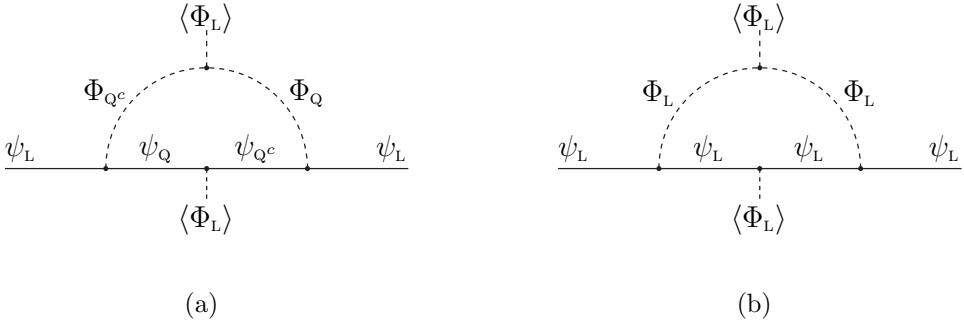
\begin{figure}[t]
  \centering
  \subfigure[]{
    \label{fig:one-loop:a}
    \centering \scalebox{0.95}{
      \begin{picture}(200,80)(-20,-10)
        \Line(0,20)(160,20)
        \Text(0,28)[l]{$\psi_\Le$}
        \Text(60,28)[]{$\psi_\qu$} \Text(100,28)[]{$\psi_{\qu^c}$}
        \Text(160,28)[r]{$\psi_\Le$}
        \DashCArc(80,20)(40,0,180){2}
        \DashLine(80,60)(80,75)2 \Text(80,80)[]{$\VEV{\Phi_\Le}$}
        \Vertex(40,20)1 \Vertex(120,20)1
        \Vertex(80,20)1 \Vertex(80,60)1
        \DashLine(80,20)(80,8)2 \Text(80,0)[]{$\VEV{\Phi_\Le}$}
        \Text(40,50)[]{$\Phi_{\qu^c}$} \Text(120,50)[]{$\Phi_\qu$}
      \end{picture}
    } }
  \subfigure[]{
    \label{fig:one-loop:b}
    \centering \scalebox{0.95}{
      \begin{picture}(200,80)(-20,-10)
        \Line(0,20)(160,20)
        \Text(0,28)[l]{$\psi_\Le$}
        \Text(60,28)[]{$\psi_\Le$} \Text(100,28)[]{$\psi_\Le$}
        \Text(160,28)[r]{$\psi_\Le$}
        \DashCArc(80,20)(40,0,180){2}
        \DashLine(80,60)(80,75)2 \Text(80,80)[]{$\VEV{\Phi_\Le}$}
        \Vertex(40,20)1 \Vertex(120,20)1
        \Vertex(80,20)1 \Vertex(80,60)1
        \DashLine(80,20)(80,8)2 \Text(80,0)[]{$\VEV{\Phi_\Le}$}
        \Text(40,50)[]{$\Phi_\Le$} \Text(120,50)[]{$\Phi_\Le$}
      \end{picture}
    } }
  \caption{{\slshape One-loop diagrams that contribute to
      neutrino masses via (a) colored Higgs and fermion fields and (b)
      color-singlet Higgs and fermion fields.  In case (a) there is
      another diagram in which the Yukawa vertices are interchanged.}
    \label{fig:one-loop}}
\end{figure}

As discussed at the end of the previous section, we can understand why
$\nu,N_1$ pair up to form an electroweak-scale Dirac neutrino, and
$N_2$ acquires an eV-scale Majorana mass, in terms of accidental
global symmetry.  If this symmetry is not respected by the Higgs
potential, however, there are large radiative contributions to these
masses.  The gauge symmetries allow the cubic Higgs couplings
$\Phi_{\qu^c} \Phi_\qu \Phi_\Le$ and $\Phi_\Le \Phi_\Le \Phi_\Le +
\text{cyclic}$, and these violate the accidental global ${\sf U(1)_X}$
symmetry of the gauge and Yukawa interactions, under which the Higgs
fields carry charge $-1$. Thus there are large (unification-scale)
radiative contributions to neutrino masses due to Higgs exchange.

In order to see the one-loop diagrams that are responsible for the
large radiative contributions to the neutrino masses, we must
consider the Yukawa interactions that are obtained from the cyclic
permutation of Eq.~(\ref{eq:yukawa-quark}).  Concentrating on just
one of the Higgs fields, we have
\begin{align} \label{eq:yukawa-quark-full}
  \mathscr{L}_q & = g \left(\psi_{\qu^c}\, \psi_\qu \Phi_\Le +
    \psi_\Le\, \psi_{\qu^c} \Phi_\qu +\psi_\qu\, \psi_\Le
    \Phi_{\qu^c}\right) + \text{h.c.}
\end{align}
These three interactions may be used to construct the one-loop diagram
in Fig.~\ref{fig:one-loop:a}.  This diagram also makes use of the
cubic Higgs coupling \mbox{$\Phi_{\qu^c} \Phi_\qu \Phi_\Le$} (with the
Higgs field $\Phi_\Le$ acquiring a unification-scale vev), as dictated
by the symmetry argument above.  A similar diagram, which instead
makes use of the Yukawa coupling \mbox{$\psi_\Le\, \psi_\Le \Phi_\Le$}
and the cubic Higgs coupling \mbox{$\Phi_\Le \Phi_\Le \Phi_\Le$}, is
shown in Fig.~\ref{fig:one-loop:b}.

The diagram in Fig.~\ref{fig:one-loop:a} is in the interaction basis.
To calculate the contribution of this diagram to the neutrino masses,
we must work in the mass-eigenstate basis. In order to perform such a
calculation, one must specify the Higgs potential, which determines
the Higgs-field mass eigenstates.  The potential has many terms, which
have been identified in Ref.~\cite{Babu:1985gi}.  Rather than pursue
such a calculation in gory detail, we idealize the situation to make
the calculation tractable, yet maintain all the qualitative features
of a full calculation.

We make the following simplifications.  First, we consider just one of
the two Higgs fields, $\Phi^1\equiv \Phi$.  Second, we consider only
the dimension-two and -three terms in the Higgs potential, and ignore
the quartic interactions.  These terms are
\begin{align} \label{eq:higgs-potential}
  \mathscr{L}_h & = m^2 \left( \Phi_\qu^\ast \Phi_\qu +
    \Phi_{\qu^c}^\ast \Phi_{\qu^c} + \Phi_\Le^\ast \Phi_\Le \right) +
  \left[ \gamma_1 \Phi_{\qu^c} \Phi_\qu \Phi_\Le +
    \gamma_2\left(\Phi_\Le \Phi_\Le \Phi_\Le + \text{cyclic} \right) +
    \text{h.c.} \right] ,
\end{align}
with $m,\gamma_i={\cal O}\left(M_\text{U}\right)$.\footnote{In
  Ref.~\cite{Babu:1985gi} it is claimed that the cubic coupling
  $\gamma_i$ must be small compared to $v$ in order to justify a
  one-loop perturbative calculation.  No such restriction appears to
  be necessary.}

We use the following notation for the colored Higgs bosons in
terms of component fields:
\begin{align} \label{eq:higgs-colored}
  \Phi_\qu & = \left(  -\mathscr{D}_\hi \ \mathscr{U}_\hi \
    \mathscr{B}_\hi \right) , &
  \Phi_{\qu^c} & =
  \begin{pmatrix}
    \mathscr{D}_\hi^c \cr \mathscr{U}_\hi^c \cr \mathscr{B}_\hi^c
  \end{pmatrix}
  .
\end{align}
The two cubic couplings, $\gamma_1 \Phi_{\qu^c} \Phi_\qu \Phi_\Le$ and
$\gamma_2 \Phi_\Le \Phi_\Le \Phi_\Le$, are given in terms of component
fields analogously to Eqs.~(\ref{eq:yukawa-quark-b}) and
(\ref{eq:yukawa-lepton2}).

We first consider the contribution to the neutrino masses from the
cubic coupling $\gamma_1 \Phi_{\qu^c} \Phi_\qu \Phi_\Le$.  We see from
Fig.~\ref{fig:one-loop:a} that the diagram is dominated by the quark
that acquires a unification-scale mass, namely the heavy $B$ quark.
Therefore the relevant scalar fields in the loop are the ${\sf
  SU(2)_L}$-singlet, down-type Higgs fields $\mathscr{D}^{c}_\hi$,
$\mathscr{B}_\hi$ and $\mathscr{B}^{c}_\hi$.  In the idealized
potential, Eq.~(\ref{eq:higgs-potential}), only $\mathscr{B}_\hi$ and
$\mathscr{B}^{c}_\hi$ mix, but not
$\mathscr{D}^{c}_\hi$.\footnote{With a more general potential, the
  mass eigenstates are linear combinations of $\mathscr{B}_\hi$,
  $\mathscr{B}^{c\ast}_\hi$, and $\mathscr{D}^{c\ast}_\hi$.}  Thus the
mass eigenstates are
\begin{subequations} \label{eq:higgs-le/ri}
  \begin{align} \label{eq:higgs-le/ri-ev}
    B_{1,2\,\hi} & = \frac{1}{\sqrt{2}} \left( \pm \mathscr{B}_\hi +
      \mathscr{B}^{c\ast}_\hi \right) , &
    B_{3\hi} & = \mathscr{D}^{c\ast}_\hi ,
    \intertext{with masses}
    \label{eq:higgs-le/ri-ew}
    m_{B_{1,2\,\hi}}^2 & = m^2 \pm \gamma_1 v_1 \ , &
    m_{B_{3\hi}}^2 & = m^2 ,
  \end{align}
\end{subequations}
where we neglect the tiny contribution from electroweak-scale
vevs.

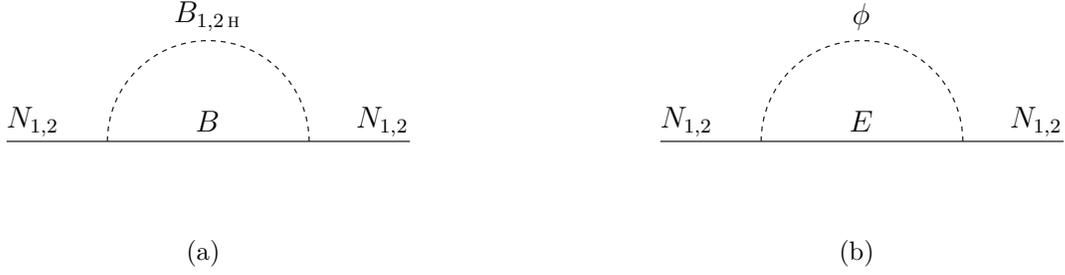
\begin{figure}[t]
  \centering
  \subfigure[]{
    \label{fig:one-loop-dominant:a}
    \centering \scalebox{0.95}{
      \begin{picture}(240,60)(-40,0)
        \Line(0,20)(160,20)  \Text(0,28)[l]{$N_{1,2}$}
        \Text(160,28)[r]{$N_{1,2}$}
        \Text(80,28)[]{$B$}  \Text(80,70)[]{$B_{1,2\,\hi}$}
        \DashCArc(80,20)(40,0,180){2}
      \end{picture}
    } }
  \subfigure[]{
    \label{fig:one-loop-dominant:b}
    \centering \scalebox{0.95}{
      \begin{picture}(240,60)(-40,0)
        \Line(0,20)(160,20)  \Text(0,28)[l]{$N_{1,2}$}
        \Text(160,28)[r]{$N_{1,2}$}
        \Text(80,28)[]{$E$}  \Text(80,70)[]{$\phi$}
        \DashCArc(80,20)(40,0,180){2}
      \end{picture}
    } }
  \caption{{\slshape Dominant one-loop diagrams, in the
      mass-eigenstate basis, that contribute to the sterile neutrino
      masses via (a) colored Higgs and fermion fields and (b)
      color-singlet Higgs and fermion fields.  In both cases there are
      actually two diagrams, which differ in the direction that
      fermion number flows in the internal fermion propagator.} 
    \label{fig:one-loop-dominant}}
\end{figure}

We now derive the vertices of the dominant contributions in terms of
these mass eigenstates (see Fig.~\ref{fig:one-loop-dominant:a}).  The
second term of Eq.~(\ref{eq:yukawa-quark-full}) gives
\begin{align} \label{eq:higgs-ri}
  \psi_\Le \psi_{\qu^c} \Phi_\qu & = \left[ -\left( E^c\, u^c +
      \mathscr{E}\, \mathscr{D}^c + \mathscr{L}\, \mathscr{B}^c
    \right) Q_\hi + \left( e^c\, u^c + \mathscr{N}_1\, \mathscr{D}^c +
      \mathscr{N}_2\, \mathscr{B}^c \right) \mathscr{B}_\hi \right] ,
\end{align}
with $Q_\hi\equiv\left(\mathscr{U}_\hi,\mathscr{D}_\hi\right)$.  The
final two terms yield the relevant neutrino interactions, written in
terms of the mass eigenstates as
\begin{align} \label{eq:higgs-ri-n}
  \psi_\Le \psi_{\qu^c} \Phi_\qu & \ni \left[ \left( s_\beta N_1 -
      c_\beta N_2 \right) \left( -s_\alpha d^c + c_\alpha B^c \right)
    - \left( c_\beta N_1 + s_\beta N_2 \right) \left( c_\alpha d^c +
      s_\alpha B^c \right) \right] \frac{1}{\sqrt{2}} \left( B_{1\hi}
    - B_{2\hi} \right) .
\end{align}
The singlet fields $N_{1,2}$ couple to $B_{1,2\,\hi}$, but $\nu$
does not, and none of the fermion fields couple to $B_{3\hi}$.
The third term of Eq.~(\ref{eq:yukawa-quark-full}) gives
\begin{align} \label{eq:higgs-le}
  \psi_\qu \psi_\Le \Phi_{\qu^c} & = \left[ \left( Q E^c + B\, e^c
    \right) U^c_\hi + \left( Q \mathscr{E} + B\, \mathscr{N}_1 \right)
    \mathscr{D}^c_\hi + \left( Q \mathscr{L} + B\, \mathscr{N}_2
    \right) \mathscr{B}^c_\hi \right] ,
\end{align}
the last two terms of which yield the neutrino interactions
\begin{align} \label{eq:higgs-le-n}
  \psi_\qu \psi_\Le \Phi_{\qu^c} \ni \left[ -d\, s_\beta\, \nu - B
    \left( c_\beta N_1 + s_\beta N_2 \right) \right]
  \frac{1}{\sqrt{2}} \left( B_{1\hi}^\ast + B_{2\hi}^\ast \right) .
\end{align}

With these vertices, we construct the one-loop contribution to the
$N_{1,2}$ two-point functions shown in
Fig.~\ref{fig:one-loop-dominant:a}.  There are actually two diagrams,
which differ in the direction that fermion number flows in the
internal fermion propagator.  The loop integral gives
\begin{align} \label{eq:integral}
  f\left(m_s,m_f\right) & = \int \frac{d^n k}{(2\pi)^4}\
  \frac{i}{k^2-m_s^2} \left[ \frac{1-\gamma_5}{2}\
    \frac{\slashed{p}+\slashed{k}+m_f}{(p+k)^2-m_f^2}\
    \frac{1-\gamma_5}{2} + \frac{1+\gamma_5}{2}\
    \frac{-\slashed{p}-\slashed{k}+m_f}{(p+k)^2-m_f^2}\
    \frac{1+\gamma_5}{2} \right] \nonumber
  \\
  & = - \frac{m_f}{(4\pi)^2} \left\{ \frac{2}{4-n} - \gamma_E +
    \log\left(4\pi\right) - \frac{m_f^2}{m_f^2-m_s^2} \log m_f^2 +
    \frac{m_s^2}{m_f^2-m_s^2} \log m_s^2 + 1 \right\} \ .
\end{align}
The integral is proportional to the fermion mass, as anticipated from
Fig.~\ref{fig:one-loop:a}.

The two-point function has contributions from both $B_{1\hi}$ and
$B_{2\hi}$, and they enter with the opposite sign (see
Eq.~(\ref{eq:higgs-ri-n})).  Thus the total contribution to the
two-point function is proportional to
\begin{align}
  \begin{split} \label{eq:Fq}
    F_B& = \frac{1}{2} \left[ f \left( m_{B_{2\hi}},m_B \right)
      - f \left( m_{B_{1\hi}},m_B \right) \vphantom{\frac12} \right]
    \\[6pt]
    & = \frac{m_B}{(4\pi)^2}\, \frac{1}{2} \left(
      \frac{m_{B_{1\hi}}^2}{m_B^2 - m_{B_{1\hi}}^2} \log
      \frac{m_{B_{1\hi}}^2}{m_B^2} - \frac{m_{B_{2\hi}}^2}{m_B^2 -
        m_{B_{2\hi}}^2} \log \frac{m_{B_{2\hi}}^2}{m_B^2} \right) .
  \end{split}
\end{align}
The ultraviolet divergences cancel, leaving a contribution to the
sterile neutrino masses proportional to the fermion mass and the Higgs
mass in the loop.

The analogous contribution to the neutrino two-point function induced
by the cubic Higgs coupling $\Phi_\Le \Phi_\Le \Phi_\Le$ is shown in
Fig.~\ref{fig:one-loop-dominant:b}.  The quark $B$ is replaced by the
lepton doublet $E$, and the Higgs fields $B_{1,2\,\hi}$ are replaced
by ${\sf SU(2)_L}$-doublet, color-singlet Higgs fields.  Recall that
in order to have successful gauge unification (in the
non-supersymmetric model), these Higgs doublet fields must lie at the
weak scale. Since the total contribution to the neutrino two-point
function, Eq.~(\ref{eq:Fq}), is proportional to the mass of the Higgs
fields in the loop, this contribution is negligible.  In the
split-SUSY scenario, this contribution would be the same order as the
contribution of Fig.~\ref{fig:one-loop-dominant:a}.

Including the large one-loop contributions, the neutrino mass
matrix, in the $\left( \nu, N_1, N_2 \right)$ basis, is
\begin{align} \label{eq:neutrino-mass}
  M_N^\text{1-loop} \simeq
  \begin{pmatrix}
    0 & -h_1 u_2 & 0 \cr -h_1 u_2 & s_{\alpha-\beta} c_\beta\, g^2 F_B
    & \left( s_{2\beta} s_\alpha - c_\alpha \right) g^2 F_B \cr 0 &
    \left( s_{2\beta} s_\alpha - c_\alpha \right) g^2 F_B &
    c_{\alpha-\beta} s_\beta\, g^2 F_B
  \end{pmatrix} ,
\end{align}
where we neglect the tiny tree-level Majorana mass of $N_2$
(Eq.~(\ref{eq:lepton-masses})).  This matrix (after factoring out
$g^2\,F_B$) is of the form \cite{Altarelli:2004za}
\begin{align} \label{eq:nu-matrix}
  \begin{pmatrix}
    0 & \epsilon & 0 \cr \epsilon & 1 & 1 \cr 0 & 1 & 1
  \end{pmatrix} ,
\end{align}
where $\epsilon \sim \frac{h_1u_2}{g^2\,F_B}$.  It has two eigenvalues
${\cal O}\left(1\right)$ and one ${\cal O}\left(\epsilon^2\right)$.
Thus the two sterile neutrinos acquire unification-scale masses at one
loop, while the active neutrino acquires a ``radiative see-saw''
Majorana mass
\begin{align} \label{eq:nu-ew}
  m_{N_{1,2}} & \sim g^2 F_B \ , &
  m_{\nu} \sim \frac{h_1^2 u_2^2}{g^2 F_B} \ .
\end{align}
In order to obtain the correct values for the tau and top masses, we
expect $h_1\simeq 0.1$, $g\simeq 1$ and \mbox{$u_2={\cal
    O}\left(10^2\,\text{GeV}\right)$}.  Since
$F_B\simeq\frac{1}{(4\pi)^2}\,M_\text{U}$, the mass of the light
neutrino is then ${\cal O}\left(0.1\,\text{eV}\right)$, consistent
with the experimental constraints \cite{nu-masses}.

There is also a one-loop diagram that couples $\nu$ to $N_{1,2}$, of
the form of Fig.~\ref{fig:one-loop:a} but with the heavy $B$ quark
replaced by $d$.  This diagram is of order
$\frac{g^2}{(4\pi)^2}\,m_d$, which could be comparable to the
tree-level Dirac mass $h_1u_2$.  In any case, it does not
qualitatively change the radiative see-saw mechanism.\footnote{It is
  claimed in Ref.~\cite{Babu:1985gi} that one should also consider the
  two-loop contribution to the $\nu$-$\nu$ two-point function obtained
  by sewing together two one-loop $\nu$-$N_{1,2}$ diagrams; however,
  such a diagram is one-particle reducible, and does not contribute to
  the mass matrix.\label{ft:two-loop}}

The radiative see-saw mechanism is absent in models with weak-scale
supersymmetry, since the one-loop contributions are reduced to ${\cal
  O}\left(1\,\text{TeV}\right)$ (due to the non-renormalization of the
superpotential in the limit of exact supersymmetry) so that the
nonvanishing entries in the neutrino mass matrix are all of the same
order.  This is analogous to the absence of the radiative seesaw
mechanism in supersymmetric {\sf SO(10)} \cite{so10-nu-mass}.  In the
supersymmetric model, one must add higher-dimensional operators or
additional Higgs representations to obtain a light, active neutrino
\cite{susy-su3}.  On the other hand, the radiative seesaw mechanism is
present if the mass difference between scalars and fermions is
comparable to the grand-unified scale, as is the case in split
supersymmetry.\footnote{This was recently discussed for {\sffamily
    SO(10)} in Ref.~\cite{Bajc:2004hr}.}  The lifetime of the gluino
restricts the sfermion masses, $m_\text{s}\lesssim
10^{14}\,\text{GeV}\simeq M_\text{U}$ \cite{Gambino:2005eh}; thus
sfermions near the upper bound yield neutrino masses in the desired
range, as in the nonsupersymmetric case.

We derived the neutrino mass matrix, Eq.~(\ref{eq:neutrino-mass}), in
a simplified model, both in the sense of the weak-scale vevs ($n_i=0$,
see Eq.~(\ref{eq:higgs})) and the Higgs potential (see
Eq.~(\ref{eq:higgs-potential})).  However, the radiative see-saw
mechanism is independent of these details, since it is governed by the
violation of the accidental global symmetry by the cubic terms in the
Higgs potential.

\newpage

\section{Three Generations \label{se:3gen}}

As we showed in Section~\ref{se:yukawa}, with one generation there is
no relation between the masses of the quarks and the charged lepton.
The same continues to be true when we extend the model to three
generations.  The Yukawa couplings $g_{1,2}$ and $h_{1,2}$ become
three-by-three matrices, and there are many more parameters than there
are constraints.  The one qualitative prediction of the model, the
radiative see-saw mechanism for the light neutrino mass, extends to
three generations, as we discuss in Section~\ref{se:neutrino-3gen}.
In the following section, we consider what the model predicts for the
CKM matrix.

\subsection{Quark Mixing \label{se:quark-mixing}}

The breaking of the electroweak symmetry leads to mixing of $d$ and
$B$, so we do not expect the CKM matrix to be exactly unitary.  Here
we show that it is unitary to a very good approximation, up to
corrections ${\cal O}\left(\frac{M_\ew}{M_\text{U}}\right)$.
However, like the quarks masses, the model does not make any
prediction for the CKM matrix.

From Eqs.~(\ref{eq:yukawa-quark}), we read off the $6\times 6$ mass
matrix of the down-type quarks for three generations,
\begin{align} \label{eq:MD-3gen}
  {\sf M}^D =
  \begin{pmatrix}
    -g_1 u_1 & g_2 v_2 \cr 0 & g_1 v_1
  \end{pmatrix} ,
\end{align}
where $g_{1,2}$ are now $3\times 3$ matrices.  ${\sf M}^D$ can be
diagonalized by unitary rotations $U_D$ and $V_D$, where\footnote{For
  the mixing among the right-handed states, see
  Appendix~\ref{se:mixing}; $V_D$ is displayed in Eq.~(\ref{eq:VD}).}
\begin{align} \label{eq:UD}
  U_D^\dagger {\sf M}^{D\dagger} {\sf M}^D
  U_D & =
  \begin{pmatrix}
    \left({\sf M}_\text{diag}^d\right)^2 & 0 \cr 0 & \left({\sf
        M}_\text{diag}^B\right)^2
  \end{pmatrix}
  , &
  U_D & =
  \begin{pmatrix}
    X & Y_1 \cr Y_2 & Z
  \end{pmatrix} .
\end{align}
Due to the unitarity of $U_D$, the four $3\times 3$ submatrices
fulfill
\begin{align} \label{eq:quark-mix}
  X X^\dagger + Y_1 Y_1^\dagger & = 1 \ , &
  X Y_2^\dagger + Y_1 Z^\dagger & = 0 \ , &
  X^\dagger Y_1 + Y_2^\dagger Z & = 0 \ , &
  Y_2 Y_2^\dagger + Z Z^\dagger & = 1 \ .
\end{align}
Combining Eqs.~(\ref{eq:MD-3gen}) and (\ref{eq:UD}) gives the
relations
\begin{subequations} \label{eq:rel}
  \begin{align} \label{eq:rel1}
    X^\dagger \left( u_1^2\, g_1^\dagger g_1\, X - u_1 v_2\,
      g_1^\dagger g_2\, Y_2 \right) + Y_2^\dagger \left( -u_1 v_2\,
      g_2^\dagger g_1\, X + v_1^2\, g_1^\dagger g_1\, Y_2 + v_2^2\,
      g_2^\dagger g_2\, Y_2 \right) & = \left({\sf
        M}_\text{diag}^d\right)^2 ,
    \\[1mm]
    \label{eq:rel2}
    X^\dagger \left( u_1^2\, g_1^\dagger g_1\, Y_1 - u_1 v_2\,
      g_1^\dagger g_2\, Z \right) + Y_2^\dagger \left( -u_1 v_2\,
      g_2^\dagger g_1\, Y_1 + v_1^2\, g_1^\dagger g_1\, Z + v_2^2\,
      g_2^\dagger g_2\, Z \right) & = 0 \ ,
    \\[1mm]
    \label{eq:rel3}
    Y_1^\dagger \left( u_1^2\, g_1^\dagger g_1\, X - u_1 v_2\,
      g_1^\dagger g_2\, Y_2 \right) + Z^\dagger \left( -u_1 v_2\,
      g_2^\dagger g_1\, X + v_1^2\, g_1^\dagger g_1\, Y_2 + v_2^2\,
      g_2^\dagger g_2\, Y_2 \right) & = 0 \ ,
    \\[1mm]
    \label{eq:rel4}
    Y_1^\dagger \left( u_1^2\, g_1 g_1^\dagger\, Y_1 - u_1 v_2\, g_1
      g_2^\dagger\, Z \right) + Z^\dagger \left( -u_1 v_2\, g_2
      g_1^\dagger\, Y_1 + v_1^2\, g_1 g_1^\dagger\, Z + v_2^2\, g_2
      g_2^\dagger\, Z \right) & = \left({\sf M}_\text{diag}^B\right)^2
    .
  \end{align}
\end{subequations}
One linear combination of Eqs.~(\ref{eq:rel}) leads to the simple
relation\footnote{The combination reads \mbox{$X
    \left[\text{Eq.}\,\left(\text{\ref{eq:rel1}}\right)\right]
    X^\dagger + X
    \left[\text{Eq.}\,\left(\text{\ref{eq:rel2}}\right)\right]
    Y_1^\dagger + Y_1
    \left[\text{Eq.}\,\left(\text{\ref{eq:rel3}}\right)\right]
    X^\dagger + Y_1
    \left[\text{Eq.}\,\left(\text{\ref{eq:rel4}}\right)\right]
    Y_1^\dagger$}.}
\begin{align} \label{eq:mixing-rel}
  X \left({\sf M}_\text{diag}^d\right)^2 X^\dagger + Y_1 \left({\sf
      M}_\text{diag}^B\right)^2 Y_1^{\dagger} = u_1^2\, g_1^\dagger
  g_1 \ ,
\end{align}
where we have used Eqs.~(\ref{eq:quark-mix}).  Since two of the three
terms in this equation are manifestly ${\cal O}\left(M_\ew^2\right)$,
we see that $Y_1$ must be ${\cal
  O}\left(\frac{M_\ew}{M_\text{U}}\right)$ in order to compensate the
${\cal O}\left(M_\text{U}\right)$ entries from ${\sf
  M}_\text{diag}^{B}$.  Thus the mixing between $d$ and $B$ is ${\cal
  O}\left(\frac{M_\ew}{M_\text{U}}\right)$.

Since $Y_1$ is very small, the first relation of
Eq.~(\ref{eq:quark-mix}) implies that the matrix $X$ is approximately
unitary; this matrix represents the generational mixing amongst the
$d$ states.  The CKM matrix is given by
\begin{align}
  U_u^\dagger X \equiv V_\text{CKM} \ .
\end{align}
where $U_u$ is the unitary matrix that diagonalizes
the up-type quark mass matrix,
\begin{align} \label{eq:up-mixing}
  U_u^\dagger \left( u_2^2\, g_1^\dagger g_1 \right) U_u = \left({\sf
      M}_\text{diag}^u\right)^2 .
\end{align}
Since $X$ is approximately unitary, the CKM matrix is unitary up to
terms \mbox{${\cal O}\left(\frac{M_\ew}{M_\text{U}}\right)$}.


\subsection{Neutrino masses \label{se:neutrino-3gen}}

In Ref.~\cite{Glashow:1984gc}, it is claimed that the light neutrino
masses would naturally have an inverted hierarchy.  This can be
understood as follows.  As seen in Section~\ref{se:neutrino}, the
eigenvalues are proportional to $\frac{h^2}{g^3}$.  Neglecting the
small mixing between the quarks, and assuming the hierarchy of $h$ is
not stronger than that of $g$, $\frac{h^2}{g^3}$ decreases from the
first to third generation yielding an inverted hierarchy.  In
contrast, in Ref.~\cite{Babu:1985gi}, the two-loop contribution (see
Footnote~\ref{ft:two-loop}) is claimed to result in a normal
hierarchy.  In this section we will show that our present
understanding of the Yukawa couplings naturally gives either
quasi-degenerate masses or a normal hierarchy.

For our discussion, we consider models with family symmetries, which
have been used extensively to study neutrino masses and mixings
\cite{Altarelli:2004za}.  A model based on ${\sf [SU(3)]^3 \times
  U(1)_F}$ was introduced in Ref.~\cite{Lola:1999un}, with the
up-quark mass matrix for three generations
\begin{align} \label{eq:up-lola}
  {\sf M}^u \sim %
  \begin{pmatrix}
    \epsilon^4 & \epsilon^3 & \epsilon^3 \cr \epsilon^3 & \epsilon^2 &
    \epsilon^2 \cr \epsilon & 1 & 1
  \end{pmatrix}
  u_2 \ , \qquad \epsilon^2 \sim \frac{m_c}{m_t} \ .
\end{align}
This choice of ${\sf M}^u$, together with a similar mass matrix for
the down quarks, leads to a viable CKM matrix.  In the following, let
us assume that $g$ (the matrix in Eq.~(\ref{eq:up-lola}), see
Eq.~(\ref{eq:quarkmasses})) is of this form.  To obtain a hierarchical
structure for the up and down quarks, both $g_1$ and $g_2$ will
generally be hierarchical.  Then, from Eq.~(\ref{eq:b-mass}), we
expect the $B$ quarks to have a similar hierarchy.

Since the one-loop contributions to the neutrino masses are
proportional to the fermion mass in the loop (see Eq.~(\ref{eq:Fq})),
those with the heaviest quark, $B_3$, are dominant.  With $g$ as given
in Eq.~(\ref{eq:up-lola}), the three-generational mass matrix for the
sterile neutrinos (both $N_1$ and $N_2$) reads (see
Fig.~\ref{fig:one-loop-dominant:a})
\begin{align} \label{eq:sterile-eff}
  {\sf M}^N \sim %
  \left( g^{3i}\, g^{j3} + g^{i3}\, g^{3j} \right) F_{B_3} \sim %
  \begin{pmatrix}
    \epsilon^4 & \epsilon^3 & \epsilon \cr \epsilon^3 & \epsilon^2 & 1
    \cr \epsilon & 1 & 1
  \end{pmatrix} F_{B_3} \ ,
\end{align}
with the eigenvalues
\begin{align} \label{eq:sterile-masses}
  m^N_3 & \sim m^N_2 \sim F_{B_3} \ , & %
  m^N_1 & \sim \epsilon^4 F_{B_3} \ ,
\end{align}
where $m^N_i$ denote the quasi-degenerate masses of both $N_1$ and
$N_2$ (of the $i$-th generation).  Since
$F_{B_3}\simeq\frac{1}{(4\pi)^2}\,M_\text{U}$, the masses of the
sterile neutrinos are $m^N_3\sim m^N_2 \sim 10^{12}\,\text{GeV}$ and
$m^N_1\sim 10^8\,\text{GeV}$.\footnote{The mass $m^N_1$ is comparable
  to the mass of the lightest sterile neutrino in thermal leptogenesis
  \cite{leptogenesis}.}

\smallskip

Let us turn to the light neutrinos.  To get a qualitative picture, we
first consider only the second and third generation.  Then we need to
find the two weak-scale eigenvalues of the effective $6\times 6$
neutrino mass matrix, given by the generalization of
Eqs.~(\ref{eq:neutrino-mass}) and (\ref{eq:nu-matrix}) to two
generations.  After a straightforward calculation, we find
\begin{multline} \label{eq:neutrino-masses-2gen}
  m^\nu_{2,3} \sim \frac{u_1^2}{F_{B_3}} \left\{ \left[
      \left(h^{22}\right)^2 + h^{22}h^{23} + \left(h^{23}\right)^2 +
      h^{23}h^{33} + \left(\epsilon h^{33}\right)^2 \right] \right.
  \\
  \pm \left. \sqrt{ \left[ \left(h^{22}\right)^2 + h^{22}h^{23} +
        \left(h^{23}\right)^2 + h^{23}h^{33} + \left(\epsilon
          h^{33}\right)^2 \right]^2 + \left[ h^{22} h^{33} +
        \left(h^{23}\right)^2 \right]^2 } \right\} .
\end{multline}
Hence, if $h^{23}\sim h^{33}$, the masses are almost degenerate; if
$h^{23}$ is smaller than $h^{33}$ and $h^{22}\lesssim\epsilon h^{23}$,
they are hierarchical.

This result holds for the three-generational case as well.  The
eigenvalues are only proportional to $\frac{h^2}{g^2}$ due to the
common loop-integral, where $g^2$ is given by the third column of the
symmetric matrix in Eq.~(\ref{eq:sterile-eff}).  Since this hierarchy
is weak, the neutrino hierarchy is determined by the hierarchy of $h$
and we find either quasi-degenerate masses or a normal hierarchy.


\section{Proton Decay \label{se:pd}}

\begin{figure}
  \centering
  \subfigure[]{\label{fig:proton-decay-rl}
    \begin{picture}(210,50)(-40,0)
      \Text(0,10)[l]{$\psi_\qu$}
      \Text(0,50)[l]{$\psi_\qu$}
      \Line(15,10)(45,30)        \Line(15,50)(45,30)
      \DashLine(45,30)(85,30)2
      \Line(63,32)(67,28)        \Line(63,28)(67,32)
      \Text(55,25)[t]{$\Phi_\qu$}
      \Text(75,25)[t]{$\Phi_\qu^\ast$}
      \Line(85,30)(115,10)       \Line(85,30)(115,50)
      \Text(135,10)[r]{$\psi_{\qu^c}^\ast$}
      \Text(130,50)[r]{$\psi_\Le^\ast$}
    \end{picture}
  }
  \subfigure[]{\label{fig:proton-decay-fusion}
    \begin{picture}(210,60)(-40,0)
      \Text(0,10)[l]{$\psi_\qu$}
      \Text(0,50)[l]{$\psi_\qu$}
      \Line(15,10)(45,30)        \Line(15,50)(45,30)
      \DashLine(45,30)(85,30)2
      \DashLine(65,30)(65,45)2
      \Line(63,47)(67,43)        \Line(63,43)(67,47)
      \Text(65,60)[t]{\small $\VEV{\Phi_\Le}$}
      \Text(55,25)[t]{$\Phi_\qu$}
      \Text(75,25)[t]{$\Phi_{\qu^c}$}
      \Line(85,30)(115,10)       \Line(85,30)(115,50)
      \Text(130,10)[r]{$\psi_\qu$}
      \Text(130,50)[r]{$\psi_\Le$}
    \end{picture}
  }
  \caption{\slshape Proton decay diagrams corresponding to the
    operators (a) $\psi_\qu\psi_\qu\psi_{\qu^c}^\ast\psi_\Le^\ast$ and
    (b) $\psi_\qu\psi_\qu\psi_\qu\psi_\Le$.
    \label{fig:proton-decay}}
\end{figure}
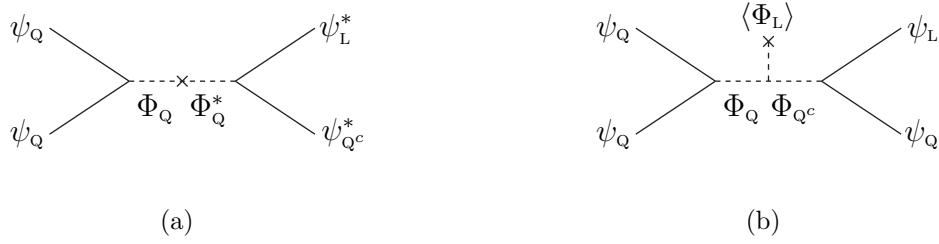

The gauge interactions conserve baryon number, and therefore do not
mediate proton decay.  Let us consider the Yukawa interactions.  The
Yukawa coupling that generates quark masses,
Eq.~(\ref{eq:yukawa-quark}), plus its cyclic permutations are
displayed in Eq.~(\ref{eq:yukawa-quark-full}),
\begin{align} \label{eq:yukawa-quark-full-prime}
  \mathscr{L}_q & = g \left(\psi_{\qu^c}\, \psi_\qu \Phi_\Le + \psi_\Le\,
    \psi_{\qu^c} \Phi_\qu +\psi_\qu\, \psi_\Le \Phi_{\qu^c}\right) +
  \text{h.c.}
\end{align}
If we assign baryon number 0 to $\Phi_\Le$, $\frac{1}{3}$ to
$\Phi_\qu$, and $-\frac{1}{3}$ to $\Phi_{\qu^c}$, then this Yukawa
interaction conserves baryon number.  However, consider the Yukawa
interaction that generates lepton masses,
Eq.~(\ref{eq:yukawa-lepton}), plus its cyclic permutations:
\begin{align} \label{eq:yukawa-lepton-full-prime}
  \mathscr{L}_l & = h \left(\psi_\Le \psi_\Le \Phi_\Le + \psi_\qu
    \psi_\qu \Phi_\qu +\psi_{\qu^c} \psi_{\qu^c}\Phi_{\qu^c}\right) +
  \text{h.c.}
\end{align}
In this case baryon number is conserved if we assign baryon number 0
to $\Phi_\Le$, $-\frac{2}{3}$ to $\Phi_\qu$, and $\frac{2}{3}$ to
$\Phi_{\qu^c}$.  Thus, with both Yukawa interactions present, baryon
number is not conserved \cite{Glashow:1984gc}.\footnote{With both
  Yukawa interactions present, only a $\mathbbm{Z}_3$ subgroup of
  baryon number is respected, where the fermion fields
  $(\psi_\Le,\psi_\qu,\psi_{\qu^c})$ and Higgs fields
  $(\Phi_\Le,\Phi_\qu,\Phi_{\qu^c})$ carry charges $(0,1,2)$.} The
amplitude for proton decay is therefore proportional to $g\,h$.
Proton-decay diagrams are generated either by putting together two
interactions containing the same Higgs field, such as $\psi_\Le
\psi_{\qu^c} \Phi_\qu$ and $\psi_\qu \psi_\qu \Phi_\qu$
(Fig.~\ref{fig:proton-decay-rl}), or two interactions containing
different Higgs fields, connected via the cubic term
$\Phi_{\qu^c}\Phi_\qu\Phi_\Le$ in the potential (see
Eq.~(\ref{eq:higgs-potential})), with $\Phi_\Le$ acquiring a
unification-scale vev (Fig.~\ref{fig:proton-decay-fusion}).  The Higgs
field $\Phi_\Le$ does not mediate proton decay because it carries the
same baryon number in both the quark and leptonic Yukawa interactions.
Thus only the colored Higgs fields, $\Phi_\qu$ and $\Phi_{\qu^c}$,
mediate proton decay.

As discussed in the Introduction, the analysis in
Ref.~\cite{Babu:1985gi} has two shortcomings. First, the authors use
the technique for a three-body decay.  Second, they use only one of
the couplings in Eq.~(\ref{eq:yukawa-lepton}), so the leptonic Yukawa
couplings are diagonal.  As a consequence the decay channel
$p\to\mu^+K^0$ is absent, because a diagonal matrix $h$ forbids the
vertex $us$ present in the exchange diagram.\footnote{The annihilation
  diagram is absent because there are no Higgs particles with charges
  $\pm\frac{4}{3}$.}  With both leptonic Yukawa couplings present in
Eq.~(\ref{eq:yukawa-lepton}) the couplings are not diagonal, so the
exchange diagram is allowed.  We will show that $p\to\mu^+K^0$ is one
of the dominant decay modes.

Proton decay is described by four effective operators of dimension
six.  To derive them, consider the Yukawa couplings with the heavy
colored Higgs bosons (Eqs.~(\ref{eq:higgs-ri}), (\ref{eq:higgs-le}))
and recall that only $\mathscr{B}_\hi$ and $\mathscr{B}^c_\hi$ mix
with each other (Eq.~(\ref{eq:higgs-le/ri})) when using the idealized
potential of Eq.~(\ref{eq:higgs-potential}).  Thus the relevant
couplings for proton decay in Eq.~(\ref{eq:yukawa-quark-full-prime})
are
\begin{align} \label{eq:cplgs-decay-1}
  \mathscr{L}_q \ni g \left[ e^c u^c \frac{1}{\sqrt{2}} \left(
      B_{1\hi} - B_{2\hi} \right) + Q\, \hat s_\beta L\,
    \frac{1}{\sqrt{2}} \left( B_{1\hi}^\ast + B_{2\hi}^\ast \right)
  \right] + \text{h.c.} ,
\end{align}
where we use the mass eigenstates and $\hat s_\beta$ denotes the
three-generational analogue of the mixing between $\mathscr{E}$ and
$\mathscr{L}$ (see Eqs.~(\ref{eq:lepton-mixing}) in
Appendix~\ref{se:mixing}).  Similarly in
Eq.~(\ref{eq:yukawa-lepton-full-prime}), where
\begin{align} \label{eq:yukawa-lepton-full-detail}
  \begin{split}
    \psi_\qu \psi_\qu \Phi_\qu & = Q Q \mathscr{B}_\hi + B Q Q_\hi \ ,
    \\
    \psi_{\qu^c} \psi_{\qu^c}\Phi_{\qu^c} & = \mathscr{D}^c u^c
    \mathscr{B}^c_\hi + u^c \mathscr{B}^c \mathscr{D}^c_\hi +
    \mathscr{B}^c \mathscr{D}^c \mathscr{U}^c_\hi \ ,
  \end{split}
\end{align}
the relevant couplings are
\begin{align} \label{eq:cplgs-decay-2}
  \mathscr{L}_l \ni h \left[ Q Q \frac{1}{\sqrt{2}} \left( B_{1\hi} -
      B_{2\hi} \right) + d^c \left( -\hat s_\alpha^\top \right) u^c\,
    \frac{1}{\sqrt{2}} \left( B_{1\hi}^\ast + B_{2\hi}^\ast \right)
  \right] + \text{h.c.}  ,
\end{align}
with $\hat s_\alpha$, the three-generational analogue of the mixing
between $\mathscr{D}^c$ and $\mathscr{B}^c$, as given in
Eqs.~(\ref{eq:down-mixing}) in Appendix~\ref{se:mixing}.  We now
integrate out the two heavy scalars and obtain the effective operators
\begin{multline} \label{eq:proton-decay-ops}
  \mathscr{L}_\text{eff} = \frac{1}{\gamma_1^2 v_1^2 - m^4} \left[
    \gamma_1 v_1 \left( \left(g\,\hat s_\beta \right)^{ij} h^{mn}\,
      Q_m Q_n Q_i L_j + g^{ij} \left( -\hat s_\alpha^\top\,h
      \right)^{mn}\, d^c_m u^c_n e^c_i u^c_j \right)
    \vphantom{\frac12} \right.
  \\[-3pt]
  - \left. m^2 \left( g^{\ast ij} h^{mn}\, Q_m Q_n e^{c\ast}_i
      u^{c\ast}_j + \left( g\,\hat s_\beta \right)^{ij} \left( -\hat
        s_\alpha^\top\,h \right)^{\ast mn}\, d^{c\ast}_m u^{c\ast}_n
      Q_i L_k \right) \vphantom{\frac12} \right] + \text{h.c.}
\end{multline}
where we have used
\begin{align}
  \frac{1}{m_{B_{1\hi}}^2} + \frac{1}{m_{B_{2\hi}}^2} & = - \frac{2
    m^2}{\gamma_1^2 v_1^2 - m^4} \ , &
  \frac{1}{m_{B_{1\hi}}^2} - \frac{1}{m_{B_{2\hi}}^2} & = \frac{2
    \gamma_1 v_1}{\gamma_1^2 v_1^2 - m^4} \ ,
\end{align}
and we now explicitly display the generation indices.  The first two
operators, involving only left or right-handed fields, are called
{\slshape LLLL} and {\slshape RRRR} operators, respectively.  They are
proportional to $\gamma_1$ because they arise from the coupling of two
different Higgs fields via the cubic interaction in the Higgs
potential, Eq.~(\ref{eq:higgs-potential}) (see
Fig.~\ref{fig:proton-decay-fusion}).  The other operators include both
left-handed and right-handed fields, and are labeled as {\slshape
  LLRR} and {\slshape RRLL}.  They are proportional to $m^2$ because
they arise from the coupling of a Higgs field to itself via the mass
term in the Higgs potential (see Fig.~\ref{fig:proton-decay-rl}).  The
quark-lepton vertex is determined by the coupling $g$ while the
quark-quark vertex is given by $h$.

To calculate the decay rates, we evolve the operators from
$M_\text{U}$ down to the hadronic scale, $M_\text{had}$.  The ratio of
the Wilson coefficients at $M_\text{U}$ and $M_\text{had}$ is
described by a factor $A={\cal O}\left(1\right)$
\cite{Emmanuel-Costa:2003pu,Dutta:2004zh}.  At $M_\text{had}$, we
switch to the hadronic level with the aid of chiral perturbation
theory \cite{Claudson:1981gh}.  The hadron matrix elements $\langle
\text{\slshape PS} \left| \mathscr{O} \right| p \rangle$, which
describe the transition of the proton to a pseudoscalar meson via the
three-quark operator $\mathscr{O}$, depend on two coefficients in the
chiral Lagrangian, namely $\alpha$ for the {\slshape LLRR} and
{\slshape RRLL} operators, and $\beta$ for the {\slshape LLLL} and
{\slshape RRRR} operators.  These coefficients describe the transition
of the proton to the vacuum via the operator $\mathscr{O}$, and are
calculated by means of lattice QCD, yielding
\mbox{$\left|\alpha\right|=\left|\beta\right|\simeq
  0.01\;\text{GeV}^3$} \cite{matrix-element}.

\begin{table}
  \begin{align*}
    \Gamma(p\to e_i^+ \pi^0) & = \frac{m_p}{32\pi f_\pi^2}
    \left(\frac{1+D+F}{\sqrt{2}}\right)^2 \left|C_{udue_i}\right|^2
    \\[3pt]
    \Gamma(p\to \bar\nu_i \pi^+) & = \frac{m_p}{32\pi f_\pi^2}
    \left(1+D+F\right)^2 \left|C_{udd\nu_i}\right|^2
    \\[3pt]
    \Gamma(p\to e_i^+ K^0) & = \frac{m_p}{32\pi f_\pi^2}
    \left(1-\frac{m_{K^0}^2}{m_p^2}\right)^2
    \left(1+(D-F)\frac{m_p}{m_B}\right)^2 \left|C_{usue_i}\right|^2
    \\[3pt]
    \Gamma(p\to \bar\nu_i K^+) & = \frac{m_p}{32\pi f_\pi^2}
    \left(1-\frac{m_{K^+}^2}{m_p^2}\right)^2 \left|
      \left(\frac{2}{3}D\frac{m_p}{m_B}\right) C_{usd\nu_i} +
      \left(1+\frac{D+3F}{3}\frac{m_p}{m_B}\right) C_{uds\nu_i}
    \right|^2
  \end{align*}
  \caption{{\slshape Partial widths of the proton decay channels}
    \cite{Buchmuller:2004eg}.
    \label{tb:chiral}}
\end{table}

The decay rates are given in Table~\ref{tb:chiral}.  Here $m_p$ and
$m_K$ denote the masses of proton and kaon, respectively, and
$f_\pi=131$\;MeV is the pion decay constant; $m_B=1.15$\;GeV is an
average baryon mass according to contributions from diagrams with
virtual $\Sigma$ and $\Lambda$; $D=0.80$ and $F=0.46$ are the
symmetric and antisymmetric flavor-{\sffamily SU(3)} reduced matrix
elements for the axial-vector current \cite{cabibbo03}.  The
coefficients $C$ are given by
\begin{align} \label{eq:coeff}
  C & = \mathscr{C}\ \frac{1}{\gamma_1^2 v_1^2 - m^4}\ A
    \begin{cases}
      \gamma_1 v_1\, \beta & (\text{\sl LLLL, RRRR}) \cr
      - m^2\, \alpha & (\text{\sl LLRR, RRLL})
    \end{cases}
\end{align}
with $\mathscr{C}$ as given in Table~\ref{tb:c}.  Generally, $\gamma_1
v_1$ and $m^2$ are ${\cal O}\left(M_\text{U}\right)$.  In the
following discussion we approximate $C\simeq \frac{g\,
  h}{M_\text{U}^2}$.

\begin{table}
  \centering
  \begin{tabular}{c|cccc}
    \rule[0mm]{0mm}{13pt} & {\sl LLLL} & {\sl LLRR} & {\sl RRLL} &
    {\sl RRRR} \\
    \hline
    \rule[0mm]{0mm}{13pt}$\mathscr{C}_{udue_i}$ &
    $\left( g\, \hat s_\beta \right)^{1i}\, h^{11}$ & $g^{\ast i1}
    h^{11}$ & $\left( g\, \hat s_\beta \right)^{1i} \left( -\hat
      s_\alpha^\top\, h \right)^{\ast 11}$ & $g^{i1} \left( -\hat
      s_\alpha^\top\,h \right)^{11}$ \\
    \rule[0mm]{0mm}{13pt}$\mathscr{C}_{udd\nu_i}$ & $\left( g\, \hat
      s_\beta \right)^{1i} h^{11}$ & --- & $\left( g\, \hat
      s_\beta \right)^{1i} \left( -\hat s_\alpha^\top\, h
    \right)^{\ast 11}$ & --- \\
    \rule[0mm]{0mm}{13pt}$\mathscr{C}_{usue_i}$ & $\left( g\, \hat
      s_\beta \right)^{1i} h^{12}$ & $g^{\ast i1}\, h^{12}$ &
    $\left( g\, \hat s_\beta \right)^{1i} \left( -\hat s_\alpha^\top\,
      h \right)^{\ast 12}$ & $g^{i1} \left( -\hat s_\alpha^\top\, h
    \right)^{12}$ \\
    \rule[0mm]{0mm}{13pt}$\mathscr{C}_{usd\nu_i}$ & $\left( g\, \hat
      s_\beta \right)^{1i} h^{12}$ & --- & $\left( g\, \hat
      s_\beta \right)^{1i} \left( -\hat s_\alpha^\top\, h
    \right)^{\ast 12}$ & --- \\
    \rule[0mm]{0mm}{13pt}$\mathscr{C}_{uds\nu_i}$ & $\left( g\, \hat
      s_\beta \right)^{2i} h^{11}$ & --- & $\left( g\, \hat
      s_\beta \right)^{2i} \left( -\hat s_\alpha^\top\, h
    \right)^{\ast 11}$ & ---
  \end{tabular}
  \caption{\slshape Coefficients $\mathscr{C}$ in
    Eq.~(\ref{eq:coeff}).  The index $i$ refers to the lepton
    generation.
    \label{tb:c}}
\end{table}

Let us now estimate the lifetime in the different decay modes with $g$
as given by the matrix in Eq.~(\ref{eq:up-lola}), which is displayed
again in Table~\ref{tb:decay-g}.  The experimental limits are listed
in Table~\ref{tb:decay-limit-coeff}.

\begin{table}[!b]
  \centering
  \caption{{\slshape (a) Dominant coefficients for the various decay
      channels and their current experimental limit} \cite{limit-sk};
    {\slshape (b)~matrix $g$ of our discussion} \cite{Lola:1999un}.
    \label{tb:decay}}
  \subtable[\label{tb:decay-limit-coeff}]{
    \begin{tabular}{lccp{12pt}r}
      \hline
      \rule[0mm]{0mm}{13pt}mode & dominant coeff. & $g$ & & limit
      [years] \\ 
      \hline
      \rule[0mm]{0mm}{15pt}$e^+ \pi^0$ & $g^{11}h^{11}$ &
      $\epsilon^4$ & & $5.4\times 10^{33}$ \\
      $\mu^+ \pi^0$ & $g^{12}h^{11}$ & $\epsilon^3$ & & $4.3\times
      10^{33}$ \\
      $\bar\nu \pi^+$ & $g^{13}h^{11}$ & $\epsilon^3$ & & $2.5\times
      10^{31}$ \\ 
      $e^+ K^0$ & $g^{11}h^{12}$ & $\epsilon^4$ & & $1.1\times
      10^{33}$ \\ 
      $\mu^+ K^0\quad$ & $g^{12}h^{12}$ & $\epsilon^3$ & & $1.4\times 
      10^{33}$ \\ 
      $\bar\nu K^+$ & $g^{23}h^{11}$ & $\epsilon^2$ & & $2.3\times
      10^{33}$ \\
      & $g^{13}h^{12}$ & $\epsilon^3$ & & 
    \end{tabular}
  }
  \hspace{60pt}
  \subtable[\label{tb:decay-g}]{
    \begin{tabular}{c}
      \rule[0mm]{0mm}{36pt}$g \sim
      \begin{pmatrix}
        \epsilon^4 & \epsilon^3 & \epsilon^3 \cr \epsilon^3 &
        \epsilon^2 & \epsilon^2 \cr \epsilon & 1 & 1
      \end{pmatrix}$
    \end{tabular}
  }
\end{table}

The most stringent experimental limit is on $p\to e^+ \pi^0$.  Using
the numerical values for the various particle masses and constants,
discussed above, we can derive an upper limit on the product of the
Yukawa coupling matrices,
\begin{align} \label{eq:epi}
  \tau & \simeq \left(\frac{1}{g\,h}\right)^2 \times 10^{28}\
  \text{years} \qquad \Rightarrow \quad g\,h \lesssim 10^{-3} \ .
\end{align}
The decay involves particles of the first generation only, so
\mbox{$g=g^{11}\sim\epsilon^4\sim 10^{-4}$}.  Hence the decay rate is
well below the experimental limit, regardless of the (presumably
small) value of $h^{11}$.

Since the Yukawa couplings of the second generation are larger and the
Yukawa matrices are not diagonal, we expect flavor non-diagonal decays
to be dominant, in particular $p\to\bar\nu K^+$ and $p\to\mu^+ K^0$.
The constraints on the product $g\, h$ are slightly less than that of
Eq.~(\ref{eq:epi}) for two reasons: the experimental limits are lower
(Table~\ref{tb:decay-limit-coeff}), and the kaon mass suppresses the
decay rate with respect to that of a pionic decay by a factor of about
2 (Table~\ref{tb:chiral}).  We show below that the dominant decay mode
is $p\to\bar\nu K^+$, with $p\to\mu^+ K^0$ less than or comparable to
it.

Let's start with $p \to \mu^+ K^0$.  The Yukawa couplings,
corresponding to an exchange diagram, are given by $g^{12}h^{12}$
({\slshape LLLL/RRLL}) and $g^{21}h^{12}$ ({\slshape LLRR/RRRR})
(there is no annihilation diagram, as already mentioned).  Since
\mbox{$g^{12} \sim g^{21} \sim \epsilon^3$}, the decay width is at
least two orders of magnitude larger than that of $p\to e^+ \pi^0$,
assuming $h^{12} \gtrsim h^{11}$.

Next consider $p\to \bar\nu K^+$.  The experimental constraint applies
to the sum of all neutrino species.  We see from
Table~\ref{tb:decay-g} that the couplings to the muon and tau
neutrinos are of the same size, and are much greater than the coupling
to the electron neutrino, independent of the quark generation.  The
decay into neutrinos is mediated by the {\slshape LLLL} and {\slshape
  RRLL} operators only.

The channel $p\to \bar\nu K^+$ is the only channel with two different
coefficients, viz.~$uds\nu_i$ and $usd\nu_i$ (see
Table~\ref{tb:chiral}).  For $uds\nu_i$ we obtain $g^{2i}h^{11}$,
whereas $usd\nu_i$ gives $g^{1i}h^{12}$.  The latter is the same
magnitude as the coupling for $p\to \mu^+ K^0$, so the width is
comparable to or larger than that of $p\to \mu^+ K^0$.  Thus $p\to
\bar\nu K^+$ is the dominant decay mode.  The lifetime is above the
experimental limit both because the $usd\nu_i$ coefficient is
suppressed by $g^{1i}\lesssim \epsilon^3 \sim 10^{-3}$, and because
$g^{2i}\lesssim \epsilon^2$ and \mbox{$\frac{h^{11}}{h^{33}} \sim
  \frac{m_e}{m_\tau}$} sufficiently suppress the $uds\nu_i$
coefficient.

The remaining channels are suppressed relative to the dominant ones.
For \mbox{$p \to e^+ K^0$}, we read off $g^{11}h^{12}$, one factor of
$\epsilon$ less than the coefficient for $p\to\mu^+ K^0$; the width is
therefore about two orders of magnitude smaller.  The couplings
involved in $p\to\bar\nu\pi^+$ and $p\to\mu^+\pi^0$ are given by
$g^{1i}h^{11}$ and $g^{12}h^{11}$, respectively, both
$\epsilon^3\,h^{11}$ since $g^{12}\sim g^{13}\sim \epsilon^3$.  These
coefficients are one factor of $\epsilon$ smaller than the coefficient
$g^{2i}h^{11}$ of $p\to\bar\nu K^+$.  Note that $p\to\mu^+\pi^0$ was
claimed to be one of the dominant decay modes in
Ref.~\cite{Babu:1985gi}.

The decay channels are summarized in Table~\ref{tb:decay-limit-coeff}.
If we assume that $h$ is hierarchical in order to have hierarchical
charged leptons, then we see -- even without a specific form of $h$ --
that the decay into kaons is favored.

\smallskip

Let us finally discuss proton decay in the presence of supersymmetry.
For the mixed operators, {\slshape LLRR} and {\slshape RRLL}, the
discussion remains unchanged, since they arise from D terms.  The
unification scale is increased by two orders of magnitude, and so the
lifetime by eight.  Therefore proton decay via these operators will
not even be observable in future experiments which aim to reach a
lifetime of $10^{35-36}$\,years \cite{future-exp}.

On the other hand, the {\slshape LLLL} and {\slshape RRRR} operators
stem from F terms and have mass-dimension five.  When the sfermions
are integrated out, they give rise to effective four-fermion operators
of dimension six.  Thus the operators are suppressed by
\mbox{$\left(m_s M_\text{U}\right)^2$} instead of $M_\text{U}^4$.  The
decay rate is naturally consistent with the experimental limit if the
sfermion masses, $m_s$, are above a few hundred TeV
\cite{Wells:2004di}.  Hence, the model with weak-scale SUSY needs
fine-tuning among the Yukawa couplings (which is similar to models
such as {\sffamily SU(5)} \cite{Emmanuel-Costa:2003pu}), whereas
proton decay is unobservable in the split-SUSY case.



\section{Conclusions \label{se:conclusion}}

In this paper we studied the trinified model, $\gtr$, with the minimal
Higgs sector required for symmetry breaking, namely two copies of
$\left(1,3,3^\ast\right)$ and its cyclic permutations.  After breaking
to the standard model there are five Higgs doublets, and
gauge-coupling unification results at $M_\text{U}\simeq 10^{14}$\,GeV
if all five are at the weak scale, without supersymmetry.  Baryon
number is conserved by the gauge interactions, so such a low
unification scale is not forbidden by proton decay.  Unlike other
grand-unified theories, such as {\sffamily SU(5)} or {\sffamily
  SO(10)}, the minimal model is able to correctly describe the fermion
masses and mixing angles without the need to introduce intermediate
scales, additional Higgs fields, or higher-dimensional operators.
Indeed, with a relatively low unification scale, it is plausible that
the effects of higher-dimensional operators induced by Planck-scale
physics are negligible.

Light, active neutrinos are naturally generated at one loop via the
radiative seesaw-mechanism.  The additional matter, which is either
vectorlike or sterile, is superheavy with masses above $10^8$\,GeV.
Thus no additional particles are present at the weak scale.

Proton decay is mediated by colored Higgs bosons.  We found that the
proton lifetime is above the experimental bounds in all the possible
decay modes due to the small Yukawa couplings.  The dominant decay
modes, $p\to\bar\nu K^+$ and $p\to\mu^+ K^0$, are potentially
observable in future experiments.

Minimal trinification is perhaps the simplest viable nonsupersymmetric
unified theory.  There are also viable models based on {\sffamily
  SO(10)}, but they require intermediate scales as well as non-minimal
Higgs sectors and/or higher-dimensional operators
\cite{Bajc:2005zf}.

We also considered the minimal model with supersymmetry.  The scenario
with weak-scale SUSY fails both because it requires additional Higgs
fields or higher-dimensional operators to generate viable neutrino
masses, and because fine-tuning is needed in order to avoid too rapid
proton decay.  In contrast, the radiative see-saw mechanism for
neutrino masses operates if the mass scale of the sfermions is around
$10^{14}$\,GeV, as in split-SUSY, which is near their upper bound
based on the gluino lifetime.  These large sfermion masses suppress
proton decay such that it is unobservable.

The dominant decay modes in minimal trinification are similar to those
in grand-unified models with weak-scale supersymmetry, where the decay
is dominated by dimension-five operators.  In {\sffamily SU(5)} or
{\sffamily SO(10)}, dimension-six operators mediated by gauge bosons
are suppressed but potentially observable since they yield decay into
pions; the estimated lifetime for $p\to e^+ \pi^0$ is \mbox{$10^{35\pm
    1}\,\text{years}$} \cite{Marciano:1997jb}.  Therefore the
different types of models would be distinguished by the presence or
absence of supersymmetric particles and the number of Higgs doublets
at the weak scale, together with the observation of specific proton
decay modes \cite{Wiesenfeldt:2004qa}.

The smoking gun for minimal nonsupersymmetric trinification would be
the discovery of five Higgs doublets at the weak scale, together with
the observation of proton decay into final states containing kaons.
The split-SUSY version of the theory would be difficult to prove, as
its main difference with split-SUSY {\sffamily SU(5)} or {\sffamily
  SO(10)} would be the absence of observable proton decay.  We look
forward to probing electroweak symmetry breaking at the Fermilab
Tevatron and the CERN Large Hadron Collider in the near future.

\subsubsection*{Acknowledgements}

We are grateful for valuable discussions with R.~Corrado.
This work was supported in part by the U.~S.~Department of Energy
under contract No.~DE-FG02-91ER40677.


\begin{appendix}


\section{Fermion masses in the general case \label{se:ferm-gen}}

For good measure, we present the general formulae for the fermion
masses in one generation where the five Higgs doublets all acquire
non-vanishing vevs $u_i$ and $n_i$.

Then the quark masses read up to corrections ${\cal
  O}\left(\frac{M_\ew}{M_\text{U}}\right)$
\begin{subequations} \label{eq:masses-gen}
  \begin{align} \label{eq:b-gen}
    m_B & = \sqrt{\left(g_1 v_1 + g_2 v_3\right)^2 + \left(g_2
        v_2\right)^2} \ ,
    \\
    \label{eq:up-gen}
    m_u & = g_1 u_2 + g_2 n_2 \ ,
    \\
    \label{eq:down-gen}
    m_d & = \frac{\left(g_1 u_1 + g_2 n_1\right) \left(g_1 v_1 + g_2
        v_3\right) - g_2^2 n_3 v_2}{m_B} \ ,
  \end{align}
  and those of the leptons are
  \begin{align} \label{eq:l-gen}
    m_E & = \sqrt{\left(h_1 v_1 + h_2 v_3\right)^2 + \left(h_2
        v_2\right)^2} \ ,
    \\
    \label{eq:electron-gen}
    m_e & = \frac{\left(h_1 u_1 + h_2 n_1\right) \left(h_1 v_1 + h_2
        v_3\right) - h_2^2 n_3 v_2}{m_E} \ ,
    \\
    \label{eq:neutrino-gen}
    m_{\nu, N_1} & = h_1 u_2 + h_2 n_2 \ ,
    \\
    \label{eq:n2-gen}
    m_{N_2} & = \frac{\left(h_1 u_2 + h_2 n_2\right) \left[ \left(h_1
          u_1 + h_2 n_1\right) \left( h_1 v_1 + h_2 v_3\right) - h_2^2
        v_2 n_3\right]}{m_E^2} \ .
  \end{align}
\end{subequations}
If $v_3$ is smaller than ${\cal O}\left(M_\text{U}\right)$, it can
be neglected.


\section{Global {\sffamily U(1)} Symmetries \label{se:global}}

The gauge sector of $\gtr$ has an accidental global ${\sf U(1)_Q
  \times U(1)_{Q^c} \times U(1)_L}$ symmetry, as discussed in
Section~\ref{se:model}.  This global symmetry is violated by the
Yukawa couplings of Eqs.~(\ref{eq:yukawa-quark}),
(\ref{eq:yukawa-lepton}), leaving a single ${\sf U(1)_X}$ global
symmetry
\begin{align}
  X\left(\Phi\right) & = -\,1 \ , & %
  X\left(\psi\right) & = \tfrac{1}{2} \ .
\end{align}
Cubic Higgs couplings do not respect this global symmetry. When
$\Phi^{1,2}_\text{\sc l}$ acquire their vevs, this global symmetry
is broken; however, combinations of this symmetry with one or more
of the broken diagonal generators of the gauge symmetry will
survive as new global symmetries. We explicitly orthogonalize the
global symmetry from any local ones.

If we impose only $v_1$, then the generators $X$, $\lambda_{8\text{\sf
    L}}$, and $\lambda_{8\text{\sf R}}$ are individually broken;
however, the combination $(\lambda_{8\text{\sf L}} +
\lambda_{8\text{\sf R}})/\sqrt{2}$ remains as an unbroken local
symmetry, ${\sf U(1)_{\text{\sf L}+\text{\sf R}}}$.  We then form an
orthogonal global symmetry by combining the local generator orthogonal
to ${\sf U(1)_{\text{\sf L}+\text{\sf R}}}$ with the global generator
$X$.  The (unnormalized) combination which is unbroken by $v_1$,
i.\,e. that which gives zero charge to the Higgs component which
acquires a vev, is \mbox{$X + \frac{\sqrt{3}}{4} \left(
    \lambda_{8\text{\sf R}}-\lambda_{8\text{\sf L}} \right) \equiv
  X_1$}.  This results in the following charges:
\begin{align}
  \Phi^{1,2}_\text{\sc l} & : \frac{3}{4} %
  \begin{pmatrix}
    -2 & -2 & -1 \cr -2 & -2 & -1 \cr -1 & -1 & 0
  \end{pmatrix} , &
  \psi_\text{\sc l} & : \frac{3}{4} %
  \begin{pmatrix}
    0 & 0 & 1 \cr 0 & 0 & 1 \cr 1 & 1 & 2
  \end{pmatrix} , &
  \psi_{{\text{\sc q}}^c} & : \frac{3}{4} %
  \begin{pmatrix}
    1 \cr 1 \cr 0
  \end{pmatrix} , &
  \psi_\text{\sc q} & : \frac{3}{4}\, %
  \begin{pmatrix}
    1 & 1 & 0
  \end{pmatrix} .
\end{align}
If we now add $v_2$, then this global symmetry is broken along with
$\lambda_{3\text{\sf R}}$ and ${\sf U(1)_{\text{\sf L}+\text{\sc
      R}}}$.  The combination \mbox{$\left(\lambda_{8\text{\sf R}} +
    \lambda_{8\text{\sf L}} - \sqrt{3}\lambda_{3\text{\sc
        R}}\right)/\sqrt{2}$} remains as an unbroken local symmetry,
which is proportional to hypercharge.  The orthogonal unbroken global
charge is then given by \mbox{$-\frac{3}{10} \left( \frac{\sqrt{3}}{2}
    \left( \lambda_{8\text{\sf R}} + \lambda_{8\text{\sf L}} \right) +
    \lambda_{3\text{\sf R}} \right) + X_1 $} which we can rewrite as
\mbox{$ \frac{\sqrt{3}}{10}\lambda_{8\text{\sf R}} -
  \frac{2\sqrt{3}}{5}\lambda_{8\text{\sf L}} -
  \frac{3}{10}\lambda_{3\text{\sf R}} + X \equiv X_2$}, where the
fields carry the charges
\begin{align}
  \Phi^{1,2}_\text{\sc l} & : \frac{3}{5} %
  \begin{pmatrix}
    -2 & -3 & -2 \cr -2 & -3 & -2 \cr 0 & -1 & 0
  \end{pmatrix} , &
  \psi_\text{\sc l} & : \frac{3}{10} %
  \begin{pmatrix}
    1 & -1 & 1 \cr 1 & -1 & 1 \cr 5 & 3 & 5
  \end{pmatrix} , &
  \psi_{{\text{\sc q}}^c} & : \frac{3}{10} %
  \begin{pmatrix}
    1 \cr 3 \cr 1
  \end{pmatrix} , &
  \psi_\text{\sc q} & : \frac{3}{10}\, %
  \begin{pmatrix}
    3 & 3 & -1
  \end{pmatrix} .
\end{align}
We can now add either $u_1$ or $u_2$.  Both will break
$\lambda_{3\text{\sf L}}$, hypercharge and the global symmetry $X_2$,
and we still have a preserved local symmetry, which is of course
electric charge, proportional to \mbox{$\lambda_{8\text{\sc R}} +
  \lambda_{8\text{\sf L}} - \sqrt{3} \left( \lambda_{3\text{\sf R}} +
    \lambda_{3\text{\sf L}} \right)$}.  In the $u_2$ case, we
construct the residual global symmetry \mbox{$-\frac{9}{8} \left(
    \frac{\sqrt{3}}{5} \left( \lambda_{8\text{\sf R}} +
      \lambda_{8\text{\sf L}} \right) - \frac{3}{5}\lambda_{3\text{\sf
        R}} + \lambda_{3\text{\sf L}} \right) + X_2 =
  \frac{1}{8} \left( \sqrt{3} \left( -\lambda_{8\text{\sf R}} - 5
      \lambda_{8\text{\sf L}} \right) + 3\lambda_{3\text{\sf R}} -
    9\lambda_{3\text{\sf L}} \right) + X$}.  This yields the charges:
\begin{align}
  \Phi^{1,2}_\text{\sc l} & :  \frac{3}{4} %
  \begin{pmatrix}
    -4 & -3 & -4 \cr -1 & 0 & -1 \cr 0 & 1 & 0
  \end{pmatrix} , &
  \psi_\text{\sc l} & : \frac{3}{4} %
  \begin{pmatrix}
    -2 & -1 & -2 \cr 1 & 2 & 1 \cr 2 & 3 & 2
  \end{pmatrix} , &
  \psi_{{\text{\sc q}}^c} & : \frac{3}{4} %
  \begin{pmatrix}
    1 \cr 0 \cr 1
  \end{pmatrix} , &
  \psi_\text{\sc q} & : \frac{3}{4}\, %
  \begin{pmatrix}
    3 & 0 & -1
  \end{pmatrix} .
\end{align}
Alternatively, in the $u_1$ case, \mbox{$\frac{3}{4} \left(
    \frac{\sqrt{3}}{5} \left( \lambda_{8\text{\sf R}} +
      \lambda_{8\text{\sf L}} \right) - \frac{3}{5}\lambda_{3\text{\sc
        R}} + \lambda_{3\text{\sf L}} \right) + X_2$} is the unbroken
global symmetry.  Collecting terms gives \mbox{$
  \frac{\sqrt{3}}{4}\lambda_{8\text{\sf R}} -
  \frac{\sqrt{3}}{4}\lambda_{8\text{\sf L}} -
  \frac{3}{4}\lambda_{3\text{\sf R}} + \frac{3}{4}\lambda_{3\text{\sf
      L}} + X$}. This gives an alternate set of global charges,
\begin{align}
  \Phi^{1,2}_\text{\sc l} & : \frac{3}{2} %
  \begin{pmatrix}
    0 & -1 & 0 \cr -1 & -2 & -1 \cr 0 & -1 & 0
  \end{pmatrix} , &
  \psi_\text{\sc l} & : \frac{3}{2} %
  \begin{pmatrix}
    1 & 0 & 1 \cr 0 & -1 & 0 \cr 1 & 0 & 1
  \end{pmatrix} , &
  \psi_{{\text{\sc q}}^c} & : \frac{3}{2} %
  \begin{pmatrix}
    0 \cr 1 \cr 0
  \end{pmatrix} , &
  \psi_\text{\sc q} & : \frac{3}{2}\, %
  \begin{pmatrix}
    0 & 1 & 0
  \end{pmatrix} .
\end{align}
Both $u_1$ and $u_2$ break the electroweak symmetry to ${\sf
  U(1)}_\EM$ but yield different charges for the global symmetry.
Hence, with both $u_1$ and $u_2$, the global symmetry is broken.  This
result is independent of which $\Phi_\text{\sc l}$ contains a specific
vev, since both $\Phi^{1,2}_\text{\sc l}$ transform identically under
the gauge and global symmetries, Furthermore, the vevs $n_i$, which we
have set to zero, do not alter the discussion.


\section{Mixing among Heavy and Light States
  \label{se:mixing}}

In Section~\ref{se:quark-mixing} we discussed the mixing among the
left-handed down-type quarks for three generations; we now consider
the right-handed fields as well as the mixing among the charged
leptons.

We can write the mixing among the right-handed fields similarly to
that of the left-handed fields (cf.  Eq.~(\ref{eq:UD})),
\begin{align} \label{eq:VD}
  \begin{pmatrix}
    \mathscr{D}^c \cr \mathscr{B}^c  
  \end{pmatrix} 
  & \to V_D 
  \begin{pmatrix}
    d^c \cr B^c
  \end{pmatrix}
  , \qquad V_D =
  \begin{pmatrix}
    X_\alpha & Y_{\alpha 1} \cr Y_{\alpha 2} & Z_\alpha 
  \end{pmatrix}
  .
\end{align}
where $X_\alpha $, the rotation of $\mathscr{D}^c $ into $d^c$, is
equivalent to $-\hat s_\alpha$ in Eq.~(\ref{eq:cplgs-decay-2}).  For a
single generation, the comparison with Eq.~(\ref{eq:down-es}) yields
$X_\alpha=-Z_\alpha=-\sin\alpha$ and \mbox{$Y_{\alpha 1}=Y_{\alpha
    2}=\cos\alpha$}.

${\sf M}^D$ is diagonalized by $V_D^\top {\sf M}^D U_D$ (cf.
Eq.~(\ref{eq:UD})),
\begin{align} \label{eq:down-diagonalization}
  \begin{pmatrix}
    X_\alpha^\top & Y_{\alpha 2}^\top \cr Y_{\alpha 1}^\top &
    Z_\alpha^\top
  \end{pmatrix}
  \begin{pmatrix}
    g_1 u_1 & g_2 v_2 \cr 0 & g_1 v_1
  \end{pmatrix}
  \begin{pmatrix}
    X & Y_1 \cr Y_2 & Z
  \end{pmatrix}
  & =
  \begin{pmatrix}
    {\sf M}_\text{diag}^d & 0 \cr 0 & {\sf M}_\text{diag}^B
  \end{pmatrix}
  ;
\end{align}
however, as discussed in Section~\ref{se:yukawa}, the mixing among
$\mathscr{D}^c$ and $\mathscr{B}^c$ is dominated by the
unification-scale vevs and $u_1$ can be neglected.  We are then free
to rotate $d$ and $B$ independently, $d \to U_d\,d$ and $B \to U_B B$,
with $U_d=X$ and $U_B=Z$.  In this approximation, the transformations
above reduce the mass terms to three supermassive and three
independent massless fields.

Neglecting $u_1$, we obtain from Eq.~(\ref{eq:down-diagonalization})
\begin{align} \label{eq:app-mix}
  \left( X_\alpha^\top g_2 v_2 + Y_{\alpha 2}^\top g_1 v_1 \right) U_B
  & = 0 \ , &
  \left( Y_{\alpha 1}^\top g_2 v_2 + Z_\alpha^\top g_1 v_1 \right) U_B
  & = {\sf M}_\text{diag}^B \ .
\end{align}
We must also satisfy the unitarity relations for $V_D$,
\begin{align} \label{eq:quark-mix-r}
  X_\alpha X_\alpha^\dagger + Y_{\alpha 1} Y_{\alpha 1}^\dagger & = 1
  \ , &
  X_\alpha Y_{\alpha 2}^\dagger + Y_{\alpha 1} Z_\alpha^\dagger & = 0
  \ , &
  X_\alpha ^\dagger Y_{\alpha 1}+ Y_{\alpha 2}^\dagger Z_\alpha & = 0
  \ , &
  Y_{\alpha 2} Y_{\alpha 2}^\dagger + Z_\alpha Z_\alpha^\dagger & = 1
  \ .
\end{align}
Eqs.~(\ref{eq:app-mix}) and (\ref{eq:quark-mix-r}) together yield
\begin{align}
  g_2 v_2 U_B & = Y_{\alpha 1}^\ast {\sf M}_\text{diag}^B \ , & g_1
  v_1 U_B & = Z_\alpha^\ast {\sf M}_\text{diag}^B \ , & U_B^\dagger
  \left( v_1^2 g_1^\dagger g_1 + v_2^2 g_2^\dagger g_2 \right) U_B & =
  \left({\sf M}_\text{diag}^B\right)^2 ,
\end{align}
thus $U_B$ diagonalizes \mbox{$v_1^2 g_1^\dagger g_1 + v_2^2
  g_2^\dagger g_2$}.  

It is also useful to define unitary matrices $L_\alpha$ and
$R_\alpha$ that diagonalize $g_2 g_1^{-1}$,
\begin{align} \label{eq:l_r_alpha}
  L_\alpha^\dagger g_2 g_1^{-1} R_\alpha & = \frac{v_1}{v_2}\, {\sf
    D}_\alpha \ ,
\end{align}
where ${\sf D}_\alpha$ is a diagonal matrix.  With these definitions
it can be shown that a solution to the eigenvalue problem above with
appropriate unitary relations is:
\begin{align}
  X_\alpha & = -v_1 L_\alpha^\ast R_\alpha^{\dagger} g_1 U_B
  \left({\sf M}_\text{diag}^B\right)^{-1} V_d \equiv -\hat s_\alpha\ ,
  & 
  Y_{\alpha 1}& = v_2 g_2^\ast U_B^\ast \left({\sf
      M}_\text{diag}^B\right)^{-1} , \nonumber
  \\
  Y_{\alpha 2} & = v_2 R_\alpha^\ast L_\alpha ^{\dagger} g_2 U_B
  \left({\sf M}_\text{diag}^B\right)^{-1} V_d \ , & 
  Z_\alpha & = v_1 g_1^\ast U_B^\ast \left({\sf
      M}_\text{diag}^B\right)^{-1} .
  \label{eq:down-mixing}
\end{align}
Here $V_d$ is an arbitrary unitary matrix, which accounts for the
freedom we would expect to rotate the three degenerate massless
fields.  It will be specified when the electroweak vevs are turned on.

We can perform a similar analysis of the $E$ and $L$ mixing:
\begin{align}
  \begin{pmatrix}
    \mathscr{L} \cr \mathscr{E}
  \end{pmatrix} 
  & \to U_E
  \begin{pmatrix}
    L \cr E 
  \end{pmatrix} 
  , \qquad U_E =
  \begin{pmatrix}
    X_{\beta} & Y_{\beta 1} \cr Y_{\beta 2} & Z_{\beta}
  \end{pmatrix}
  , &
  E^c & \to V_E E^c \ ,
\end{align}
where $X_\beta\equiv\hat s_\beta$ (cf. Eq.~(\ref{eq:cplgs-decay-1})).
With
\begin{align} \label{eq:u_beta}
  V_E^\dagger \left( v_1^2 h_1 h_1^\dagger + v_2^2 h_2 h_2^\dagger
  \right) V_E & = \left({\sf M}_\text{diag}^E\right)^2 , &
  R_\beta^\dagger h_1^{-1} h_2 L_\beta & = \frac{v_1}{v_2}\, {\sf
    D}_\beta \ ,
\end{align} 
we can write
\begin{align}
  X_\beta & = v_1 L_\beta^\ast R_\beta^{\dagger} h_1^{\dagger} V_E
  \left({\sf M}_\text{diag}^E\right)^{-1} U_\ell \equiv \hat s_\beta \ 
  , & 
  Y_{\beta 1} & = v_2 h_2^\top V_E^\ast \left({\sf
      M}_\text{diag}^E\right)^{-1} , \nonumber
  \\
  Y_{\beta 2} & = v_2 R_\beta^\ast L_\beta ^{\dagger} h_2^{\dagger}
  V_E \left({\sf M}_\text{diag}^E\right)^{-1} U_\ell \ , & 
  Z_\beta & = -v_1 h_1^\top V_E^\ast \left({\sf
      M}_\text{diag}^E\right)^{-1} .
  \label{eq:lepton-mixing}
\end{align}
Again, the matrix $U_\ell$ is unspecified as long as the electroweak
vevs are zero.

When the electroweak symmetry is broken, all the above equations
obtain tiny corrections \mbox{${\cal
    O}\left(\frac{M_\ew}{M_\text{U}}\right)$}, and $V_d$ and $U_\ell$
are determined.

\end{appendix}



{\small
\begin{thebibliography}{99}
\addtolength{\itemsep}{-6pt}

\bibitem{Georgi:1974sy} H.~Georgi and S.~L.~Glashow,
  Phys.\ Rev.\ Lett.\  {\bf 32}, 438 (1974).

\bibitem{Langacker:1980js} P.~Langacker,
  Phys.\ Rept.\ {\bf 72}, 185 (1981).

\bibitem{nu-masses}
  S.~M.~Bilenky, C.~Giunti, J.~A.~Grifols and E.~Masso, 
  Phys.\ Rept.\ {\bf 379}, 69 (2003);
  \\
  A.~Strumia and F.~Vissani,
  Nucl.\ Phys.\ B {\bf 726}, 294 (2005).

\bibitem{Achiman:1978rv}
  Y.~Achiman and B.~Stech,
  in {\sl Advanced Summer Institute on New Phenomena in Lepton and
    Hadron Physics}, eds.  D.~E.~C.~Fries and J.~Wess (Plenum, New
  York, 1979).

\bibitem{Glashow:1984gc}
  S.~L.~Glashow,
  in {\sl Fifth Workshop on Grand Unification}, ed.~K.~Kang, H.~Fried,
  and P.~Frampton (World Scientific, Singapore, 1984), p.~88.

\bibitem{Ginsparg:1987ee} P.~H.~Ginsparg,
  Phys.\ Lett.\ B {\bf 197}, 139 (1987).

\bibitem{Babu:1985gi} K.~S.~Babu, X.~G.~He and S.~Pakvasa,
  Phys.\ Rev.\ D {\bf 33}, 763 (1986).

\bibitem{susy-su3}
  B.~Campbell, J.~R.~Ellis, M.~K.~Gaillard, D.~V.~Nanopoulos and
  K.~A.~Olive,
  Phys.\ Lett.\ B {\bf 180}, 77 (1986);
  \\
  B.~R.~Greene, K.~H.~Kirklin, P.~J.~Miron and G.~G.~Ross,
  Nucl.\ Phys.\ B {\bf 292}, 606 (1987);
  \\
  G.~Lazarides, C.~Panagiotakopoulos and Q.~Shafi,
  Phys.\ Lett.\ B {\bf 225}, 66 (1989);
                                %
  ibid. B {\bf 315}, 325 (1993)
  [Erratum-ibid.\ B {\bf 317}, 661 (1993)];
  \\
  G.~Lazarides and C.~Panagiotakopoulos,
  Phys.\ Lett.\ B {\bf 336}, 190 (1994);
                                %
  Phys.\ Rev.\ D {\bf 51}, 2486 (1995);
  \\
  M.~Y.~Wang and E.~D.~Carlson,
  arXiv:hep-ph/9302215;
  \\
  G.~R.~Dvali and Q.~Shafi,
  Phys.\ Lett.\ B {\bf 326}, 258 (1994);
                                %
  ibid. B {\bf 339}, 241 (1994);
                                %
  ibid. B {\bf 403}, 65 (1997);
  \\
  N.~Maekawa and Q.~Shafi,
  Prog.\ Theor.\ Phys.\ {\bf 109}, 279 (2003);
  \\
  J.~Giedt,
  Mod.\ Phys.\ Lett.\ A {\bf 20}, 2369 (2005);
  \\
  T.~Roy,
  Phys.\ Rev.\ D {\bf 71}, 035010 (2005).
  
\bibitem{orbifold-su3}
  K.~S.~Choi and J.~E.~Kim,
  Phys.\ Lett.\ B {\bf 567}, 87 (2003);
  \\
  J.~E.~Kim,
  Phys.\ Lett.\ B {\bf 564}, 35 (2003);
                                %
  ibid. B {\bf 591}, 119 (2004);
  \\
  C.~D.~Carone and J.~M.~Conroy,
  Phys.\ Rev.\ D {\bf 70}, 075013 (2004);
  \\
  C.~D.~Carone,
  Phys.\ Rev.\ D {\bf 71}, 075013 (2005);
  \\
  A.~Demaria and R.~R.~Volkas,
  Phys.\ Rev.\ D {\bf 71} (2005) 105011.

\bibitem{Willenbrock:2003ca} S.~Willenbrock,
  Phys.\ Lett.\ B {\bf 561}, 130 (2003).

\bibitem{split-susy}
  N.~Arkani-Hamed and S.~Dimopoulos,
  JHEP {\bf 0506}, 073 (2005);
  \\
  G.~F.~Giudice and A.~Romanino,
  Nucl.\ Phys.\ B {\bf 699}, 65 (2004) [Erratum-ibid.\ B {\bf 706}, 65
  (2005)].

\bibitem{Witten:1979nr} E.~Witten,
  Phys.\ Lett.\ B {\bf 91}, 81 (1980).

\bibitem{su3-s3}
  E.~D.~Carlson and M.~Y.~Wang,
  arXiv:hep-ph/9211279;
  \\
  K.~Chalut, H.~Cheng, P.~H.~Frampton, K.~Stowe and T.~Yoshikawa,
  Mod.\ Phys.\ Lett.\ A {\bf 17}, 1513 (2002).

\bibitem{Glashow:1976nt} S.~L.~Glashow and S.~Weinberg,
  Phys.\ Rev.\ D {\bf 15}, 1958 (1977).

\bibitem{fcnc}
  M.~Sher and Y.~Yuan,
  Phys.\ Rev.\ D {\bf 44}, 1461 (1991);
  \\
  A.~Antaramian, L.~J.~Hall and A.~Rasin,
  Phys.\ Rev.\ Lett.\ {\bf 69}, 1871 (1992).

\bibitem{unification}
  J.~R.~Ellis, S.~Kelley and D.~V.~Nanopoulos,
  Phys.\ Lett.\ B {\bf 260}, 131 (1991);
  \\
  U.~Amaldi, W.~de Boer and H.~Furstenau,
  Phys.\ Lett.\ B {\bf 260}, 447 (1991);
  \\
  P.~Langacker and M.~Luo,
  Phys.\ Rev.\ D {\bf 44}, 817 (1991).

\bibitem{Altarelli:2004za} G.~Altarelli and F.~Feruglio,
  New J.\ Phys.\ {\bf 6}, 106 (2004).

\bibitem{so10-nu-mass}
  T.~E.~Clark, T.~K.~Kuo and N.~Nakagawa,
  Phys.\ Lett.\ B {\bf 115}, 26 (1982);
  \\
  L.~E.~Ibanez,
  Phys.\ Lett.\ B {\bf 117}, 403 (1982).

\bibitem{Bajc:2004hr} B.~Bajc and G.~Senjanovic,
  Phys.\ Lett.\ B {\bf 610}, 80 (2005).

\bibitem{Gambino:2005eh}
  P.~Gambino, G.~F.~Giudice and P.~Slavich,
  Nucl.\ Phys.\ B {\bf 726}, 35 (2005).

\bibitem{Lola:1999un} S.~Lola and G.~G.~Ross,
  Nucl.\ Phys.\ B {\bf 553}, 81 (1999).

\bibitem{leptogenesis}
  W.~Buchm\"uller, P.~Di Bari and M.~Pl\"umacher,
  Annals Phys.\  {\bf 315}, 305 (2005);
  \\
  W.~Buchmuller, R.~D.~Peccei and T.~Yanagida,
  arXiv:hep-ph/0502169;
  \\
  O.~Vives,
  arXiv:hep-ph/0512160.

\bibitem{Emmanuel-Costa:2003pu}
  D. Emmanuel-Costa and S.~Wiesenfeldt,
  Nucl.\ Phys.\ B {\bf 661}, 62 (2003).

\bibitem{Dutta:2004zh}
  B.~Dutta, Y.~Mimura and R.~N.~Mohapatra,
  Phys.\ Rev.\ Lett.\ {\bf 94}, 091804 (2005).

\bibitem{Claudson:1981gh} M.~Claudson, M.~B.~Wise and L.~J.~Hall,
  Nucl.\ Phys.\ B {\bf 195}, 297 (1982).

\bibitem{matrix-element}
  S.~Aoki {\it et al.}  [JLQCD Collaboration],
  Phys.\ Rev.\ D {\bf 62}, 014506 (2000);
  \\
  Y.~Aoki [RBC Collaboration],
  Nucl.\ Phys.\ Proc.\ Suppl.\ {\bf 119}, 380 (2003);
  \\
  N.~Tsutsui {\it et al.}  [CP-PACS Collaboration],
  Nucl.\ Phys.\ Proc.\ Suppl.\  {\bf 129}, 284 (2004);
                                %
  Phys.\ Rev.\ D {\bf 70}, 111501 (2004).

\bibitem{Buchmuller:2004eg} W.~Buchm\"uller, L.~Covi,
  D.~Emmanuel-Costa and S.~Wiesenfeldt,
  JHEP {\bf 0409}, 004 (2004).

\bibitem{cabibbo03} N.~Cabibbo, E.~C.~Swallow and R.~Winston,
  Phys.\ Rev.\ Lett.\ {\bf 92}, 251803 (2004).

\bibitem{limit-sk}
  M.~Shiozawa [Super-Kamiokande Collaboration],
  in: {\sl Proceedings of the 28th International Cosmic Ray
    Conference} (ed. T. Kajita), Univ. Acad. Pr., 2004.
  \\
  K.~Kobayashi {\it et al.}  [Super-Kamiokande Collaboration],
  Phys.\ Rev.\ D {\bf 72}, 052007 (2005).

\bibitem{future-exp} 
  J.~G.~Learned, {\sl 9th International Symposium on Neutrino
    Telescopes, Venice, Italy, 6-9 Mar 2001}, {\tt
    http://axpd24.pd.infn.it/conference2001/proceedings/Learned.ps};
  \\
  C.~Yanagisawa,
  PoS (AHEP2003) 062;
  \\
  R.~J.~Wilkes,
  arXiv:hep-ex/0507097.

\bibitem{Wells:2004di} see, e.\,g., %
  J.~D.~Wells,
  Phys.\ Rev.\ D {\bf 71}, 015013 (2005).

\bibitem{Bajc:2005zf}
  B.~Bajc, A.~Melfo, G.~Senjanovic and F.~Vissani,
  arXiv:hep-ph/0510139.
  
\bibitem{Marciano:1997jb}
  W.~J.~Marciano,
  AIP Conf.\ Proc.\  {\bf 397}, 11 (1997).

\bibitem{Wiesenfeldt:2004qa} see, e.\,g., %
  S.~Wiesenfeldt,
  Mod.\ Phys.\ Lett.\ A {\bf 19}, 2155 (2004).

\end{thebibliography}
}
\end{document}